\documentclass{article}
\pdfoutput=1
\usepackage[dvipsnames]{xcolor}
\usepackage{jheppub}
\usepackage{lineno}
\usepackage{style}
\usepackage{braket}	
\usepackage{tikz}
\usepackage{pgfplots}
\usepackage{bm}
\usetikzlibrary{decorations.markings}
\usetikzlibrary{decorations.pathmorphing}
\usetikzlibrary{arrows.meta}
\usepackage{graphicx}
\usepackage{mathrsfs}
\usepackage{cancel}
\usepackage{subcaption}
\usepackage{float}
\usepackage{booktabs} 
\usepackage{array}
\usepackage{comment}
\usepackage{slashed}

\newcommand{\ncancel}[1]{\bcancel{#1}}

\newcommand{\nN}[1]{}

\def\nn{\nonumber}

\def\d{\mathrm{d}}

\def\ba{\begin{eqnarray}}
\def\ea{\end{eqnarray}}
\def\O{\mathcal{O}}

\def\L{\mathcal{L}}

\def\del{\nabla}

\def\d{\mathrm{d}}

\def\({\left(}
\def\){\right)}

\def\p{\partial}
\def\ie{{\em i.e. }}

\def\O{\mathcal{O}}

\def\L{\mathcal{L}}

\def\del{\nabla}

\def\nn{\nonumber}

\def\d{\mathrm{d}}

\def\({\left(}
\def\){\right)}
\def\ie{{\it i.e. }}

\def\nn{\nonumber}
\def\p{\partial}

\title{\boldmath 
I want to Break Free: The Positive Side of Symmetry Breaking}

\author[a]{Anubha Bal,}
\author[a,b,c]{Claudia de Rham}

\affiliation[a]{Abdus Salam Centre for Theoretical Physics, Imperial College, London, SW7 2AZ, UK}
\affiliation[b]{Perimeter Institute for Theoretical Physics, 31 Caroline St N, Waterloo, ON, N2L 2Y5, Canada}
\affiliation[c]{CERCA, Department of Physics, Case Western Reserve University, 10900 Euclid Ave, Cleveland, OH 44106, USA}

\emailAdd{anubha.bal21@imperial.ac.uk, c.de-rham@imperial.ac.uk}

    \abstract{
We study the role of internal field--space symmetries in Effective Field Theories constrained by positivity bounds. For theories with $N$ real scalar fields, we classify derivative operators according to the internal symmetries they preserve, organising them into hierarchies based on the maximal symmetry $O(N)$ and its subgroups. We find that the Wilson coefficients of operators that depart from an enhanced symmetry are bounded in terms of those that respect it. This means that departures from an enhanced internal symmetry cannot be made arbitrarily large while maintaining a standard UV completion, even when the enhanced internal symmetry was not imposed in constructing the low--energy theory. 
Moreover, along a cascade of progressively smaller preserved subgroups, the coefficients of operators preserving lower levels of symmetry are bounded by those preserving larger subgroups in the cascade, and are consequently subject to progressively stronger constraints. We demonstrate these results through a variety of examples involving $2-2$ and $3-3$  scattering of indefinite states formed from superpositions of multiple scalar species. }

\begin{document}
\maketitle
\flushbottom

\section{Introduction}
\label{sec:Intro}

Symmetries are at the heart of nature. They underpin the laws of physics and the models used to describe particle physics, gravity, and cosmology. In a bottom--up approach, a low--energy Effective Field Theory (EFT) is constructed by specifying a set of light fields and assumed symmetries. These assumptions determine the allowed classes of operators, which are then organised according to their relevance in the EFT expansion. As we shall see, the S-matrix bootstrap programme subsequently constrains the Wilson coefficients of these operators using general principles governing the high-energy completion of the EFT.\\

Known UV completions often exhibit a high degree of symmetry. The question we shall ask is whether a stronger, previously unassumed symmetry can instead emerge from the consistency requirements of the UV completion (irrespective of its assumed level of symmetry other than Poincar\'e invariance). In other words, can the S-matrix bootstrap reveal an enhanced symmetry that was not imposed from the outset when constructing the EFT? Our main finding is that, quite generically, UV consistency privileges low-energy EFTs with enhanced internal symmetries by bounding the size of departures from them. Thus, to be very clear, our methods are not relying on the idea that symmetry-violating operators may be added arbitrarily to an EFT whose symmetry has already been specified. Rather, an initially more general EFT may be forced by UV consistency to approach a more symmetric limit. These results generalise the findings of \cite{Remmen:2020vts,Bonnefoy:2020yee,Sun:2025axx}, which uncovered bounds on departures from flavour conservation (see also \cite{Davighi:2023acq} for related questions).\\

The S-Matrix bootstrap programme has been revitalised in modern physics, since its birth in the 1960s \cite{Martin:1965jj,Jin:1964zz,Froissart:1961ux, Hearn:1962zz, Pham:1985cr, Goldberg:1968zza, SMatrixbook, PhysRev.166.1768}. EFTs have become a critical tool in modern physics, enabling us to describe low-energy physics while remaining agnostic about the fundamental UV physics. Assuming the principles of unitarity, Lorentz invariance, locality and causality, we can constrain the space of EFTs which admit a standard (Wilsonian) UV completion. These are imposed using the S-Matrix analyticity framework, which leads to positivity bounds, imposing constraints on EFT (Wilson) coefficients with scalar, vector, and spin-$2$ field content \cite{Adams:2006sv, Jenkins:2006ia, Bellazzini:2014waa, Keltner:2015xda, Bonifacio:2016wcb, Cheung:2016yqr, Bellazzini:2016xrt, Bai:2016pae, Cheung:2016wjt, deRham:2017xox, deRham:2017zjm, deRham:2017imi, deRham:2017avq, deRham:2018qqo, Alberte:2019zhd, Alberte:2019xfh, Alberte:2020bdz, Alberte:2020jsk, Kennedy:2020ehn, Wang:2020jxr, Zhang:2020jyn, deRham:2021fpu, Noumi:2021uuv, Caron-Huot:2021rmr, Chiang:2021ziz, Li:2021lpe, deRham:2022sdl, Haring:2022sdp, Herrero-Valea:2022lfd, CarrilloGonzalez:2023cbf, Chen:2023bhu, Wan:2024eto, Xu:2024iao, Ye:2024rzr, Beadle:2024hqg, Berman:2024wyt, Lee:2025kgs, Ferretti:2026iij, Dong:2026rbc}.
These constraints can rule out regions of the Wilson coefficient parameter space that may have seemed viable when constructing the EFT. While the S-matrix bootstrap programme has already proven to be remarkably powerful, making progress with generic EFTs is challenging when the number of couplings is overwhelmingly large. A good example is the Standard Model EFT (SMEFT), which already contains over 40,000 independent Wilson coefficients \cite{Murphy:2020rsh, Li:2020gnx} at dimension-8 operator order. To best guide our searches and priors in this vast landscape, positivity bounds have been shown to constrain the volume of the SMEFT parameter space down to 0.7\% \cite{Remmen:2020vts,Bonnefoy:2020yee,Chala:2021wpj, Zhang:2018shp, Remmen:2019cyz}, while \cite{Chang:2026ztn} recently restricted anomalous quartic gauge couplings down to 0.0313\%.\\
These bounds typically make use of linear and non-linear (fully crossing-symmetric/null) constraints \cite{Tolley:2020gtv,Caron-Huot:2020cmc,Sinha:2020win} and other non-linear constraints \cite{Arkani-Hamed:2020blm,Bellazzini:2020cot},
as well as indefinite states built as superpositions of field states. In most situations, the scales probed are considered to be large enough that all fields can be treated as massless. However, the framework for deriving positivity bounds for scattering amplitudes involving indefinite states built out of non-zero and unequal mass states was derived in \cite{deRham:2025htd}, where the applicability of the bounds was extended to these situations.\\

Searches for new physics, for example beyond the Standard Model, may illuminate new symmetries of nature. While symmetry provides a powerful organising principle for this operator expansion, if one remains agnostic about the internal field space structure, and no internal global symmetry is imposed by hand, then the coefficients of all local and Lorentz invariant (in flat space, or more generally diffeomorphism invariant) operators are, in general, independent parameters of the low-energy theory. Understanding the extent to which positivity bounds constrain different classes of operators is therefore an important question for EFTs, and this shall be our main focus. \\

There is an important question as to whether an initially general low-energy EFT can depart arbitrarily far from an enhanced internal symmetry limit, by making the Wilson coefficients of the enhanced symmetry-breaking operators arbitrarily large. We will show, through several detailed examples and guiding principles, that this is not possible. The coefficients of enhanced symmetry-breaking operators are bounded in terms of those of enhanced symmetry-preserving operators. In this sense, the UV completion constrains the extent to which the low-energy theory can depart from an enhanced-symmetry phase. A consistent and standard UV completion therefore generically favours EFTs that remain as symmetric as possible, even when that enhanced symmetry was not imposed as a premise in constructing the low-energy theory.\\

\begin{figure}[h!]
    \centering
    \includegraphics[width=0.9\linewidth]{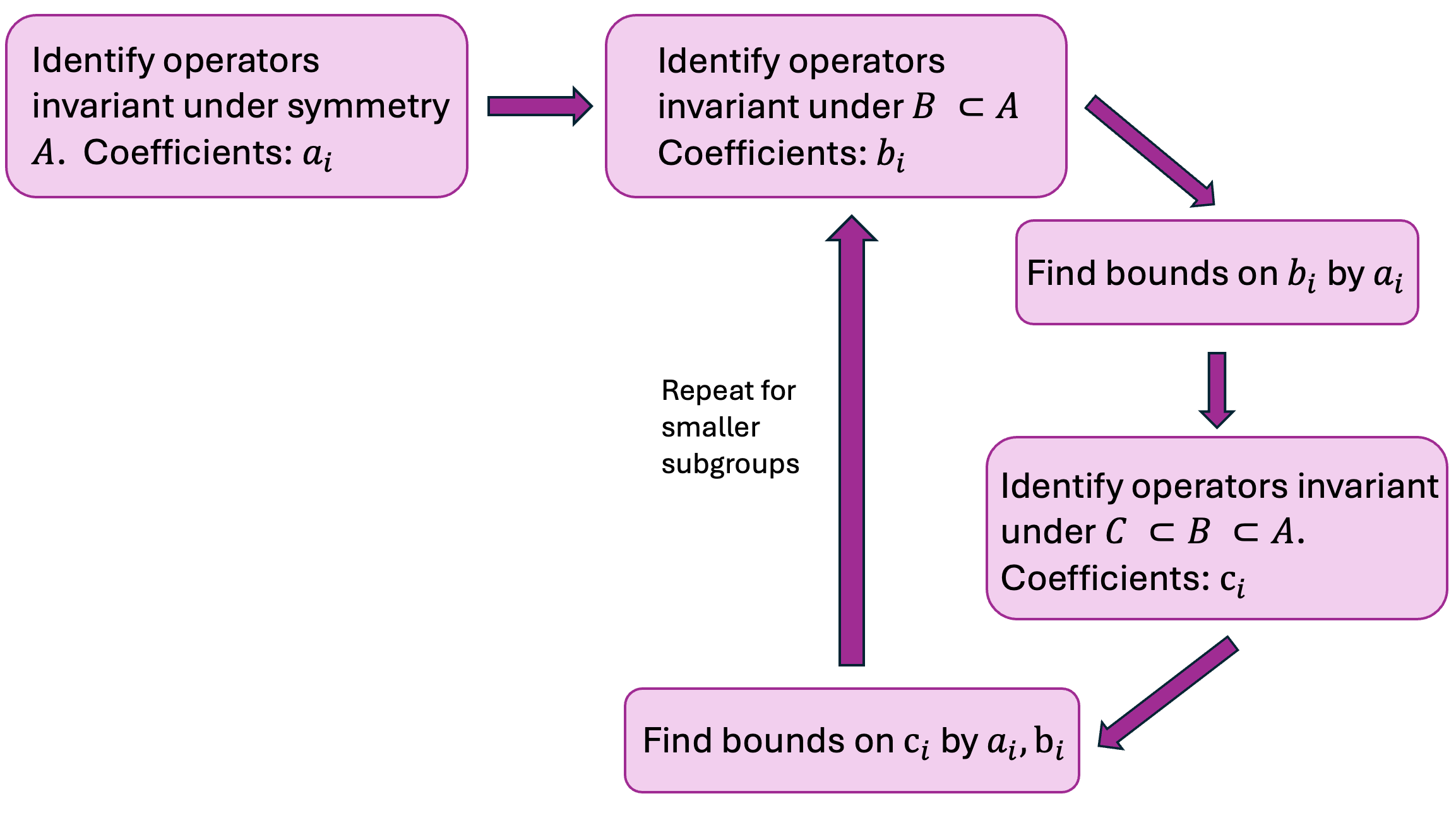}
    \caption{A flow diagram illustrating the steps we take to bound the coefficients of all of the operators for a given symmetry breaking sequence.}
    \label{fig:workflow}
\end{figure}

For concreteness, we will study $N$ real scalar fields. The highest level of internal symmetry such multi-field EFTs can satisfy is  $O(N)$.  Starting with an EFT that contains all operators, we classify the operators based on the level of symmetry they preserve, starting with those preserving the maximal $O(N)$ symmetry. We then further build our operator basis by identifying operators enjoying symmetries which are subgroups of $O(N)$, and we will show that the Wilson coefficients of these operators are bounded by those of the symmetry-preserving operators. To solidify our argument, we will proceed to classify operators in a hierarchy, delineated by the number of internal subgroups of the maximal symmetry group they preserve, and show that as we explicitly break the maximally-enhanced symmetry to smaller and smaller subgroups, the Wilson coefficients preserving each subgroup are bounded by the coefficients that preserved the larger subgroups. We find that the operators which break the most symmetries of the EFT (those at the bottom of the hierarchy) are the most tightly constrained. The bottom of the symmetry-breaking hierarchy corresponds to the most restricted directions in the EFT parameter space. This provides a natural way of organising the allowed EFT parameter space according to the degree of symmetry breaking. We note that for the EFTs considered here and at the orders we work with to impose positivity bounds for $2-2$ and $3-3$ scattering, we cannot distinguish between $O(N)$ and $SO(N)$ symmetric operators (there are no operators that are $SO(N)$ symmetric but not $O(N)$). An $O(N)$ symmetric operator is also guaranteed to have $N$ copies of $\mathbb{Z}_2$ symmetries. We also extend our approach to situations that include exceptional symmetry groups such as the compact exceptional Lie group $G_2$. In string theory and M-theory, such exceptional symmetry groups appear naturally from the compactification of extra dimensions in situations that admit a four-dimensional $\mathcal{N}=1$ supersymmetric reduction. We see here that, more generally, such symmetry groups play a critical role, as dictated by the consistency of the UV completion but independently of its precise details.
A summary of our approach can be seen in Fig.~\ref{fig:workflow}.\\

To identify operators that preserve a subgroup $B$ but break a larger symmetry group $A$, we will make sure to isolate the genuinely $A$-breaking part using the standard group-averaging projector. This is because any operator which preserves $B$ may naturally include a piece that preserves $A$. Specifically, for a candidate operator $\mathcal{O}$, we compute its $A$--invariant projection $P_A\mathcal{O}$ and define
\mbox{$\mathcal{O}^{\ncancel{A}} \equiv \mathcal{O}-P_A\mathcal{O},$}
averaged over the group parameter. 
By construction, $\mathcal{O}^{\ncancel{A}}$ contains no component invariant under $A$, while retaining the transformation properties under the subgroup $B$. This ensures that the coefficients multiplying these operators parameterise genuine departures from the $A$ symmetric, rather than shifts of the $A$-symmetric Wilson coefficients. We then apply the positivity bounds to the resulting EFT parameterised in terms of our operator basis. Most optimal bounds are derived analytically but also include numerical methods to visualise the compact bounds. We have further  devised a method to recover analytic expressions for bounds on a single coefficient using symbolic regression as given in Appendix~\ref{App:PySR}, though our fully analytic bounds are more optimal.

\paragraph{Conventions, notation and terminology:}
Throughout this paper, we work in four-dimensional Minkowski spacetime with mainly plus signature. 
We consider scalar EFTs and focus on the leading-order interactions that enter the second $s$-derivative of either the $2-2$ or $3-3$ amplitude (those are studied separately). The generalisation to theories with field states of different masses can be done as prescribed in \cite{deRham:2025htd}. When dealing with $2-2$ scattering, we work with Mandelstam variables
\begin{eqnarray} \label{Mandelstams}
s = - (k_1 + k_2)^2 \ , \qquad t = - (k_1 + k_3)^2 \qquad \mathrm{and} \qquad u = -  (k_1 + k_4)^2 \, .
\end{eqnarray}
The kinematics of $n-n$ scattering are discussed further in Appendix~\ref{App:Crossing}. We denote the symmetry-preserving coefficients of the maximal global symmetry by $a_i$. This is always $O(N)$. The Wilson coefficients corresponding to the largest subgroup of the maximal group in the hierarchy are denoted by $b_i$. We denote the coefficients corresponding to the next subgroup in the hierarchy by $c_i$, and so forth.

To derive our positivity bounds (inferred from Unitarity, Causality and locality in the Froissart-bound sense of the UV completion), we will use superposition states to calculate the amplitude, and apply positivity bounds to this. This process is discussed in detail in Appendix~\ref{App:2-2}. These mixed states contain indefinite angles and phases which we denote by $\theta_i, \varphi_i$. For example, the $2-2$ elastic amplitude can be written as 
\begin{equation}
\langle \mathbf{k}_3^A, \mathbf{k}_4^B | \hat{T} | \mathbf{k}_1^A, \mathbf{k}_2^B \rangle := (2\pi)^4 \delta^{(4)}(\mathbf{k}_1 + \mathbf{k}_2 + \mathbf{k}_3 + \mathbf{k}_4) 
\cdot \mathcal{A}_{AB \to AB},
\label{2-2Amp}
\end{equation}
where $A,B$ label the external superposition states, and $\hat{T}$ is the non-trivial part of the usual scattering operator $\hat{S}$. We then use the positivity statement 
\begin{equation}
    \partial_{s}^2\mathcal{A}_{AB \to AB}(\varphi_i,\theta_i,a_i,b_i,c_i...)>0, \quad \forall \ \varphi_i,\theta_i\,,
\label{eq:Bound}
\end{equation}
and this constraint must be satisfied for all possible superpositions, hence all values of the mixing angles $\theta_i, \varphi_i$. 
For $3-3$ indefinite scattering, we follow the framework established in \cite{Chandrasekaran:2018qmx}, and reviewed in Appendix~\ref{App:Crossing}. \vspace{0.5cm}

The rest of this paper is organised as follows: We  perform an extensive study of the bounds on symmetry breaking sequences involving continuous subgroups of the original symmetry in Section~\ref{sec:Continuous}. We begin with an EFT enjoying $O(N)$ as the maximal symmetry, and present the constraints on the Wilson coefficients of $O(N-1)$ preserving operators, which break the $O(N)$ symmetry. We then study cascades of symmetry breaking, for instance  with the sequences $O(8) \to U(4) \to \mathbb{Z}_2^4$, and $O(7) \to G_2 \to \mathbb{Z}_2^6$. In Section~\ref{sec:Discrete}, we examine the consequences of breaking a global symmetry to a discrete subgroup, and compare bounds on the Wilson coefficients of operators which break the original symmetry to varying degrees. We provide some future directions in Section~\ref{sec:Conclusion}. To complement our derivations, technical details of the dispersion relation and the indefinite bounds are provided in Appendix~\ref{App:2-2}. We also explain the procedure behind the group averaging projector in Appendix~\ref{App:GAP}. Details of the extremisation procedure and the exact optimal analytic bounds for some situations is given in Appendix~\ref{App:O8Bound}. The procedure is extended to $3-3$ scattering in Appendix~\ref{App:Crossing} following the setup exposed in \cite{Chandrasekaran:2018qmx}. Appendix~\ref{App:PySR} provides the details of the symbolic regression for analytic bounds that are not exposed in the main text but provide a consistent alternative derivation that agrees with the analytic and numerical bounds provided here. 

\section{Operator Bases and Continuous Internal--Symmetry Hierarchies}
\label{sec:Continuous}

Throughout this work, we consider a set of $N\in\mathbb{N}^*$ real Lorentz invariant scalar fields $\Phi=\{\phi^a\}_{a=1,\cdots,N}$. We work in the mass eigenstate basis with canonical kinetic terms and diagonal mass matrix. We define the internal-space tensor $ X^{ab}$
\ba
  X^{ab}=-\partial_{\mu}\phi^a \partial^{\mu}\phi^b, \quad a,b=1, ... , N \,   ,
\ea
the Lagrangian density can always be written as
\ba
\L=\frac 12 [X] -\frac12 \sum_{a=1}^N m_a^2 \phi_a^2+\L_{\rm int}(\Phi)\,,
\ea
where square brackets denote traces over internal field-space indices,  with the field-space metric $\delta_{ab}$. 
In principle, the interaction Lagrangian can contain an arbitrary number of fields and derivatives, organised according to the usual EFT expansion.

Since the aim of this work is to identify constraints from positivity bounds, we shall focus on the leading-order irrelevant operators that contribute to  the linear bounds. For bounds arising from  $2-2$ scattering, those are  the following dimension-8 operators\footnote{Naturally the theory may contain other dimension--8 operators but those do not enter the linear $2-2$ bounds at tree level.} quadratic in $X^{ab}$
\ba
\label{eq:generic interactions8}
\L_{\rm int}^{\text{dim}-8}(\Phi)=\frac{\lambda_{abcd}}{\Lambda^4}  X^{ab}X^{cd}\,,
\ea
where $\Lambda$ is the strong coupling scale. 
For  $3-3$ amplitudes, we shall assume that these leading dimension-12 operators vanish (for reasons beyond the scope of this work) and consider instead the analogous dimension-12 operators cubic in $X^{ab}$
\ba
\label{eq:generic interactions12}
\L_{\rm int}^{\text{dim}-12}(\Phi)=\frac{\lambda_{abcdef}}{\Lambda^{8}}  X^{ab}X^{cd}X^{ef}\,.
\ea
In four spacetime dimensions, the numbers of independent operators of these two forms are
\ba
N^{\text{dim}-8}&=&\frac{1}{2^3}N(N+1)(N(N+1)+2)\,,\\
{\rm and}\qquad 
N^{\text{dim}-12}&=&\frac{1}{2^33!}N(N+1)(N(N+1)+2)(N(N+1)+4)\,.
\ea
 Clearly the number of independent operators to keep track of soon becomes intractable without the help of a deeper guiding principle.  Rather than working with the (subset of) coefficients $\lambda_{abcd}$ and $\lambda_{abcdef}$ as the independent parameters, we classify the corresponding operators according to the internal symmetries they preserve,
\ba
\label{eq:Lint}
\L_{\rm int}(\Phi)=
\L^{O(N)}(\Phi)+\L^{\ncancel{O(N)},\ O(N-1)}+\L^{\ncancel{O(N)},\ \ncancel{O(N-1)},\ {\rm subgroup}}+\cdots\,,
\ea
where the ellipses denote terms preserving proper subgroups lower in the chosen symmetry-breaking hierarchy. 

To make the procedure explicit, we first construct in what follows the maximally symmetric $O(N)$-invariant operator bases and then build our basis through the smaller subgroups, first focusing on continuous subgroups and then in Section~\ref{sec:Discrete}, moving onto discrete subgroups. For a specific number of dimensions, we shall encounter exceptional symmetry groups along the way which we treat separately. 

Throughout this work we shall consider scattering processes that involve all families of mixing included in  \eqref{eq:generic interactions8} or \eqref{eq:generic interactions12}. 
 The relevant derivations are given in  Appendix~\ref{App:2-2} for $2-2$ amplitudes. For $3-3$ scattering, we follow the framework established in \cite{Chandrasekaran:2018qmx}. In both cases, we consider the scattering of indefinite states, leading to a superposition of amplitudes with subtleties discussed in Appendix~\ref{App:Crossing}. 

For simplicity, we make all the fields massless, however the generalisation of the indefinite bounds to theories with field states of different masses can be done as prescribed in \cite{deRham:2025htd}. Interestingly corrections from unequal mass states scale as $\Delta m^2/\Lambda^2$ if one assumes a tree-level weakly coupled UV completion. This implies that 
although unequal masses can break the $O(N)$ symmetry of the quadratic (free) theory by order one, their effects on positivity bounds are suppressed if one assumes a tree-level UV completion and the bounds we will derive will remain essentially immune to that quadratic order breaking. If we remain even more agnostic on the details of the UV realisation and push the analysis to allow for more generic classes of UV completion, including potentially strongly coupled ones, we see from \cite{deRham:2025htd} that the impact on the indefinite positivity bounds we shall derive are tractable and can remain parametrically weak. 

\subsection{$O(N)$ operators}

We start by formulating the set of operators that are $O(N)$ symmetric. Any scalar  trace of $X^n$ with the $\delta_{ab}$ internal field space metric is naturally manifestly $O(N)$ symmetric, though we will find it convininent to express those in terms of the following invariants
\ba
\mathcal{O}^{O(N),n}\sim \varepsilon_{a_1\cdots a_n \bullet \cdots \bullet}\varepsilon_{b_1\cdots b_n}{\!\!}^{\bullet \cdots \bullet} X^{a_1 b_1}\cdots X^{a_n b_n}\,,
\ea
where internal field space indices are raised and lowered using $\delta_{ab}$ and where $\varepsilon_{\bullet \cdots \bullet}$ is the Levi-Civita symbol. 
Using these, the relevant  $O(N)$ invariant dimension-8  operators that enter the first term in \eqref{eq:Lint} are
\ba
\label{eq:4pt}
\L^{O(N)}(\Phi)=\sum_{i=1}^2 a_i \O_{i, \ \text{dim}-8}^{O(N)} =\frac{a_1}{\Lambda^4}[X\nN{_N}]^2+\frac{a_2}{\Lambda^4}([X\nN{_N}^2]-[X\nN{_N}]^2).
\ea
For operators at sixth order in the field, the express the relevant  $O(N)$ invariant dimension-12  operators that enter the first term in \eqref{eq:Lint} as
\begin{equation}
\begin{aligned}
\L^{O(N)}(\Phi)&=\sum_{i=1}^3 a_i \mathcal{O}_{i, \ \text{dim}-12}^{O(N)} \\
    &=  \frac{a_1}{\Lambda^8}([X\nN{_N}]^3 + \frac{2}{3}[X\nN{_N}]\big([X\nN{_N}^2]-[X\nN{_N}]^2\big))\\
    &+ \frac{a_2}{\Lambda^8}[X\nN{_N}]\big([X\nN{_N}^2]-[X\nN{_N}]^2\big)
    + \frac{a_3}{\Lambda^8}\big([X\nN{_N}]^3-3[X\nN{_N}][X\nN{_N}^2]+2[X\nN{_N}^3]\big).
\end{aligned}
\label{eq:6pt}
\end{equation}
Note that when considering operators at sixth order in the fields that contribute to the $3-3$ amplitude, we always assume the absence of contamination from operators that contribute to the lower point amplitude. The generalisation to more generic EFTs will be considered in upcoming work following the approaches of \cite{Cheung:2025nhw} and \cite{Kremminger:2026edr}.
For an EFT with only two fields, the three possible combinations of traces of $X$ are not independent, and so we can set $a_3=0$ without loss of generality.

\subsection{Symmetry breaking Operator Bases}

We first identify the operators invariant under global symmetry $A$, and then identify the independent operators that are invariant under a continuous subgroup $B$ of $A$, but which break $A$. In this way, we can find bounds on operators that describe departures from the $A$-symmetry while preserving $B$. We then repeat this procedure by identifying independent operators that are invariant under a smaller subgroup $C\subset B$ but not under $A$ nor $B$, see Fig.~\ref{fig:workflow}. \\

When we break the global symmetry of an EFT by adding one or more operators, it is possible that those operators contain contributions that do enjoy the global symmetry we intended to break. To find the truly symmetry-breaking directions in the space of available independent operators, we must project out the parts of each symmetry-breaking operator that are invariant under the original symmetry. A systematic way to do this is to use the group averaging projector. Suppose we have an EFT with global symmetry $A$ with group elements $g$, parametrised by traces of $X_{ab}$, given by 
\begin{equation}
    \mathcal{L}_A=\sum_{i}a_i \mathcal{O}_i^{A}(X)\,,
\end{equation}
where the sum runs over all independent invariant operators for the EFT order we are working with. We then want to explicitly break $A$ with an operator $\mathcal{O}(X)$. We can decompose the operator as 
\begin{equation}
\mathcal{O}(X)=\mathcal{O}_0(X)+\mathcal{O}^{\ncancel{A}}(X)\,,
\end{equation}
where $\mathcal{O}_0(X)$ is invariant under $A$ and $\mathcal{O}^{\ncancel{A}}(X)$ is the genuinely symmetry breaking piece. We can write 
\begin{equation}
P_A \mathcal{O}(X)= \int_A \text{d}g\mathcal{O}(gXg^T)\,,
\end{equation}
where the measure is normalised to have weight one. We can then calculate 
\begin{equation}
    \mathcal{O}^{\ncancel{A}}=\mathcal{O}(X)-P_{A}\mathcal{O}(X)\,,
\end{equation}
as the isolated symmetry breaking operator. This is how we will calculate the basis for symmetry breaking operators in all cases.\

\subsection{Two-dimensional field space}

As an example, we consider an EFT of two real scalar fields and enumerate all independent dimension-eight operators that contribute to tree-level $2\to2$ scattering at this order. For $N=2$, the relevant symmetry hierarchy is
\ba
O(2)\supset \mathbb{Z}_2\times\mathbb{Z}_2 \supset \mathbb{Z}_2\,.
\ea
At this order, the EFT contains six independent operators. Two of them are $O(2)$-invariant, as introduced in \eqref{eq:4pt}, and have Wilson coefficients $a_{1,2}$. The remaining four operators break $O(2)$ and can be classified according to the residual symmetry they preserve. Two of these remaining four preserve the subgroup $\mathbb{Z}_2\times\mathbb{Z}_2$ and have Wilson coefficients $b_{1,2}$, while the final two preserve only the diagonal $\mathbb{Z}_2$ subgroup and have Wilson coefficients $c_{1,2}$.

\begin{figure}[h!]
    \centering
    \includegraphics[width=0.8\linewidth]{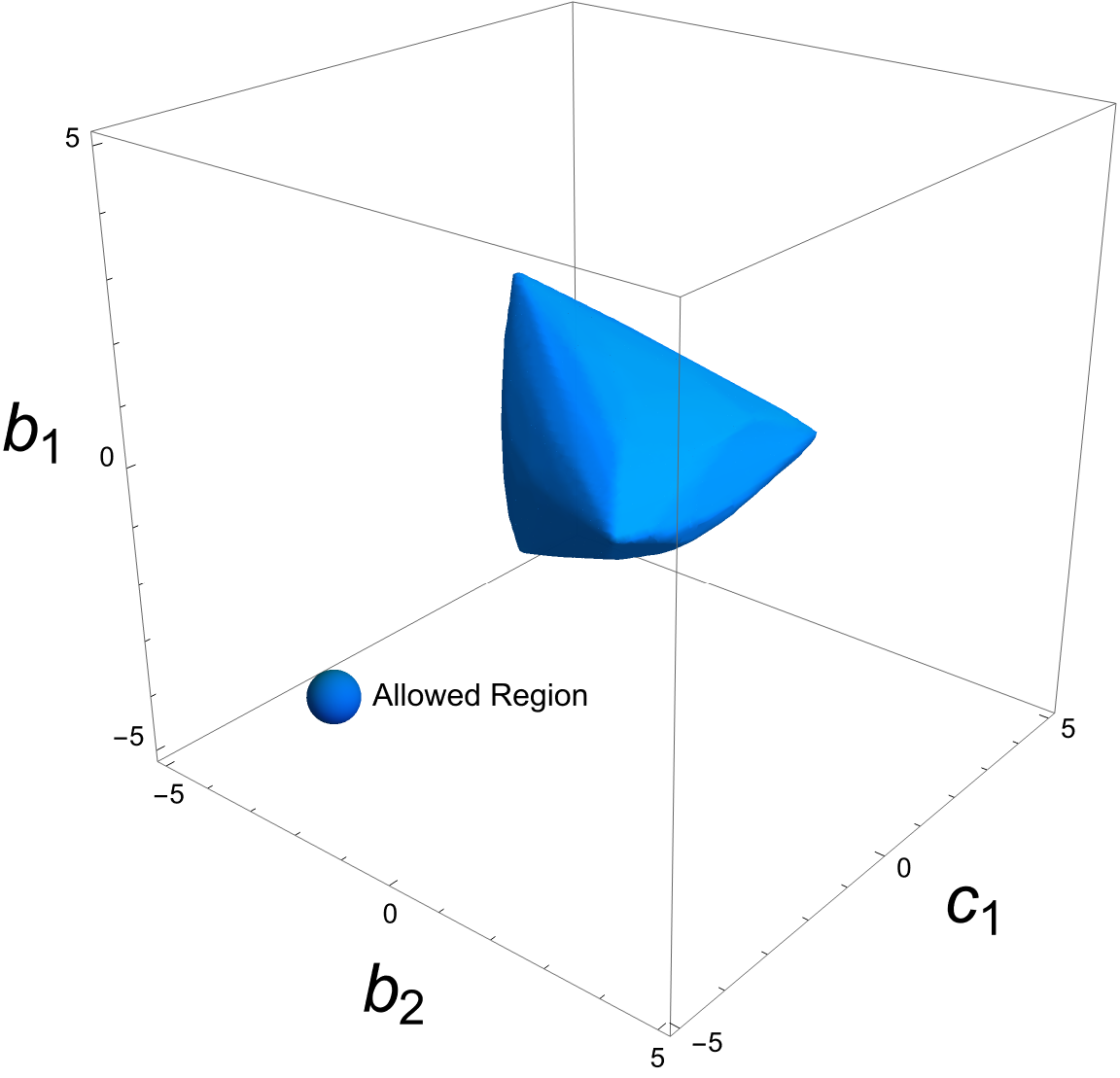}
    \caption{The two-field case considered in \eqref{eq:2fieldEFT} with all operators switched on, here depicted with specific choice of parameters $a_1=2a_2=1$ and $c_1=c_2$.}
    \label{fig:2fieldnote}
\end{figure}

To illustrate the projection procedure, we consider the $O(2)$ breaking but $\mathbb{Z}_2\times\mathbb{Z}_2$ preserving operator $X_{11}^2$ and evaluate its projection onto $O(2)$, $\mathcal{P}_{O(2)}X_{11}^2$.
The elements of $O(2)$ are two-dimensional orthogonal matrices with determinant $+1$ or $-1$. We denote the determinant $+1$ elements, which are rotations parametrised by a continuous angle $\theta$, by $R(\theta)$, and the determinant $-1$ elements by $\widetilde{R}(\theta)$. Averaging $X_{11}^2$ over the two connected components of $O(2)$ gives
\begin{equation}
\begin{aligned}
    \mathcal{P}_{O(2)}X_{11}^2&=\frac{1}{4\pi}\int_{O(2)}\ \text{d}\theta[((RXR^T)_{11})^2+((\tilde{R} X \tilde{R}^T)_{11})^2] \\ &= \frac{3}{4}(X_{11}^2+X_{22}^2)+\frac{1}{2}X_{11}X_{22}+X_{12}^2 \\&=\mathcal{P}_{O(2)}X_{22}^2\,,
\end{aligned}
\end{equation}
which implies that the operator 
\begin{equation}
    \O_1^{\ncancel{O(2)},\ \mathbb{Z}_2 \times \mathbb{Z}_2}=\frac{1}{2\Lambda^4}(X_{11}^2-X_{22}^2)
\end{equation}
is $O(2)$--breaking and contains no invariant pieces. Similarly, we consider the other independent operator that breaks $O(2)$ and preserves $\mathbb{Z}_2\times\mathbb{Z}_2$. We could either start with $X_{11}X_{22}$ or with $X_{12}^2$ so those are not independent once their projection onto $O(2)$ is properly taken into account
\begin{equation}
    X_{12}^2-\mathcal{P}_{O(2)} X_{12}^2=X_{11}X_{22}-\mathcal{P}_{O(2)} X_{11}X_{22}\,.
\end{equation}
We can therefore take the second independent $\mathbb{Z}_2 \times \mathbb{Z}_2$ operator to be
\begin{equation}
\begin{aligned}
    \O_2^{\ncancel{O(2)},\ \mathbb{Z}_2 \times \mathbb{Z}_2}&= \frac{1}{\Lambda^4}( X_{12}^2-\mathcal{P}_{O(2)} X_{12}^2) \\&=\frac{1}{\Lambda^4} \left( -\frac{1}{8}(X_{11}^2+X_{22}^2) + \frac{1}{2}X_{12}^2 + \frac{1}{4}X_{11}X_{22} \right)\,.
\end{aligned}
\end{equation}
In addition, there are two operators which preserve one $\mathbb{Z}_2$ symmetry, given by $X_{11}X_{12}$ and $X_{22}X_{12}$, which already have group averaging projectors of zero. All dimension-8 operators in this EFT are contained in
\begin{equation}
\L^{N=2}_{\rm dim-8}=\sum_{i=1}^2 a_i \O_{i}^{O(2)}+\sum_{i=1}^2 b_i \O_{i}^{\ncancel{O(2)},\ \mathbb{Z}_2 \times \mathbb{Z}_2}+\sum_{i=1}^2 c_i \O_{i}^{{\ncancel{O(2)}},\ \mathbb{Z}_2}\,.
\label{eq:2fieldEFT}
\end{equation}
Standard definite positivity bounds impose $a_{1,2}>0$. We find that the Wilson coefficients of the symmetry breaking operators (\ie the $b_i$ and $c_i$ coefficients) are bounded on both sides by the symmetry preserving coefficients $a_i$. 
When only one symmetry--breaking operator is switched on at a time, the analytic bounds are:
\begin{equation} 
\begin{array}{l l l} 
a_1,\;a_2 \ge 0 & & \text{bound always valid} \\[20pt] 
|b_1| \le 
\begin{cases} 
2a_1 & a_1\le a_2\\ 
2\sqrt{a_2(2a_1-a_2)} & a_1>a_2 
\end{cases} 
& & \text{bound valid when } b_2=c_{1,2}=0 \\[30pt] 
|b_2| \le 4\min\{a_2,2a_1\} & & \text{bound valid when } b_1=c_{1,2}=0 \\[25pt] 
|c_i|\le 
\begin{cases} 
\dfrac{16a_1}{3\sqrt3}, & a_1\le 3a_2/4,\\[6pt] 
4a_2\sqrt{1-\dfrac{a_2}{3a_1}}, & a_1>3a_2/4 
\end{cases} 
& & \text{bound valid when } b_{1,2}=c_{k\ne i}=0 \,.
\end{array} 
\label{eq:2fieldsbounds2}
\end{equation}
As indicated above, for simplicity these bounds are only given in special cases where only a few operators are tuned on at any given time. We can however explore the full parameter space, with all six operators switched on simultaneously, using a combination of analytic and numerical positivity bounds. As a separate consistency check, we also use symbolic regression to obtain analytic approximations to the resulting bounds. As an example, we consider the EFT in \eqref{eq:2fieldEFT} with all six operators included and represent the resulting bounds in Fig.~\ref{fig:2fieldnote}.

As is clear from this figure, the bounds become stronger as more symmetry-breaking operators are switched on, leading to a fully compact allowed region in the non-maximally symmetric parameter space. We will see that this pattern is generic to larger internal field spaces and for other symmetry groups. It also echoes earlier observations made in namely \cite{Alberte:2019xfh,Alberte:2019zhd} that the inclusion of additional interactions generally leads to more restrictive bounds. This finding is significant for the SMEFT, as it suggests that bounds obtained by switching on only a finite subset of operators represent the most optimistic constraint. Including additional operators is expected to further restrict the allowed parameter space rather than enlarge it.

\begin{figure}[h!]
    \centering
    \includegraphics[width=0.6\linewidth]{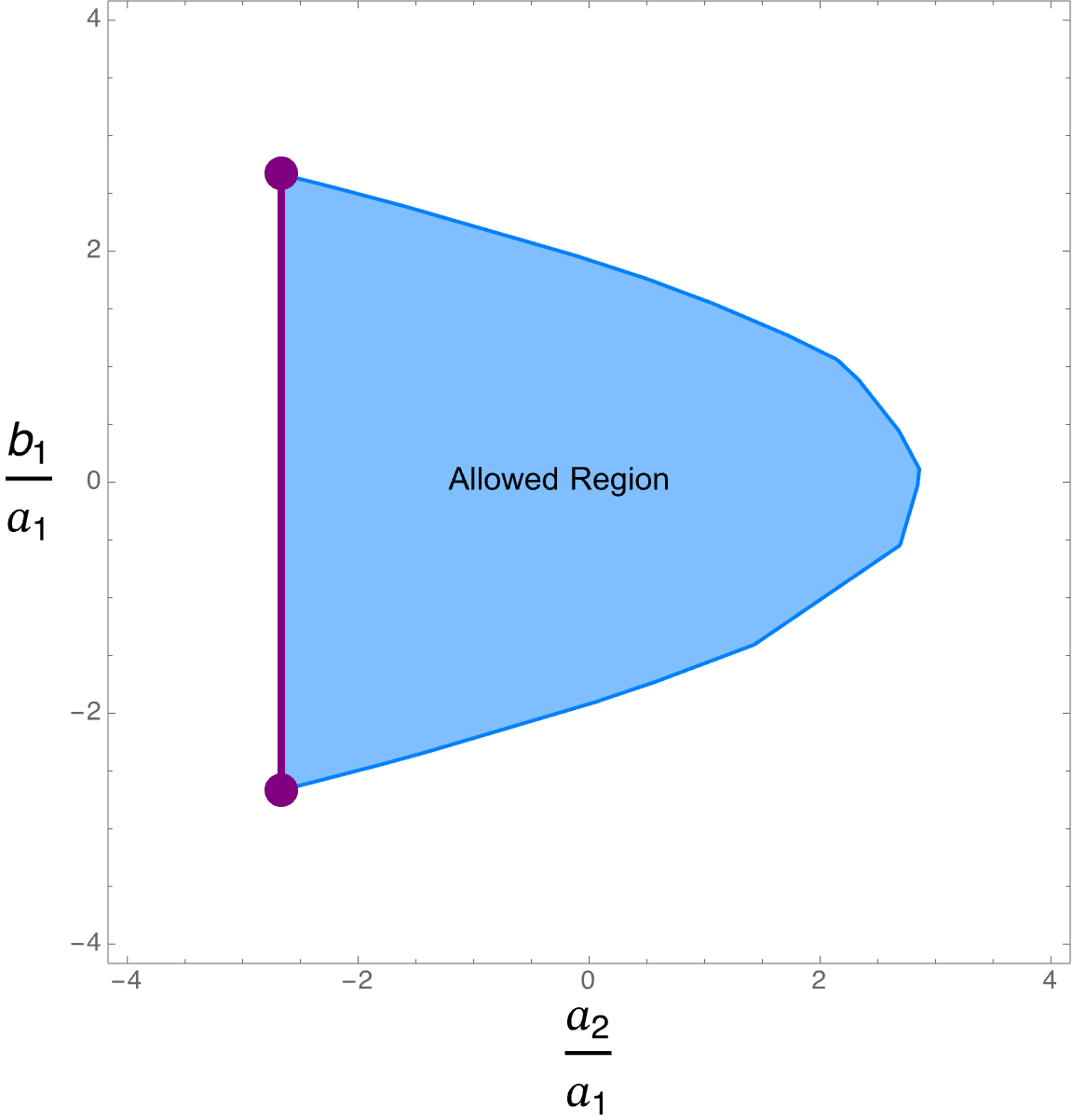}
    \caption{Two-field theory with $c_1=c_2=0$ and $a_1=\frac32 a_2=1$. The thick purple line corresponds to the choice of parameter that captures the situation where the two fields are fully decoupled, as  in \eqref{eq:decoupled}.
      Its endpoints (top and bottom purple dots) correspond to the cases $\lambda_1=a_1,\lambda_2=0$ and $\lambda_1=0,\lambda_2=a_1$, respectively, while interior points describe generic choices with $\lambda_1,\lambda_2>0$.  Any point in the purple line is allowed and captures a generic mixture of $\lambda_{1,2}>0$. Any point away from that vertical axis at $b_2=-8a_1/3$ includes mixing between both fields.}
    \label{fig:Decoupled_choice}
\end{figure}

\paragraph{Decoupled Sectors:}
The operator basis used in \eqref{eq:Lint} and in the previous example implicitly assumes that all fields may interact. 
If one field decouples completely, or if the $N$ fields split into several decoupled sectors, it would instead be natural to choose an operator basis adapted to this decomposition from the outset. Each decoupled sector would then satisfy its own independent set of positivity bounds. Nevertheless, it is instructive to ask how our results and our choice of basis adapts for situations that includes decoupled sectors.  

As a simple example, consider the two fields $\phi_1$ and $\phi_2$ that are entirely decoupled.
In that case, 
\ba
\label{eq:decoupled}
\L_{\rm int}(\Phi)=\frac{\lambda_1}{\Lambda^4} (\p \phi_1)^4+\frac{\lambda_2}{\Lambda^4} (\p \phi_2)^4=
\frac{\lambda_1}{\Lambda^4}X_{11}^2+\frac{\lambda_2}{\Lambda^4} X_{22}^2\,.
\ea
In the operator basis used so far, this corresponds to setting $c_{1,2}=0$ and imposing
\ba
a_1=\frac{3}{8}(\lambda_1+\lambda_2)= \frac 32 a_2  = -\frac 3 8 b_2\,, \quad{\rm and }\quad  b_1=\lambda_1-\lambda_2\,.
\ea
We show the slice of parameter space corresponding to $a_2=2a_1/3$ and $c_1=c_2=0$ in Fig.~\ref{fig:Decoupled_choice}. 
As is clearly visible in that figure, the limit in which the two sectors are completely decoupled is fully captured by our analysis and corresponds to a special boundary of the allowed parameter space (thick line on the right of Fig.~\ref{fig:Decoupled_choice}). Once a small interaction mixing the two fields is switched on (corresponding to $b_2>-8 a_1/3$), one moves away (to the right) of this boundary and the allowed range of $b_1$ shrinks.

\subsection{$O(N) \to O(N-1)$}
We now generalise the $O(2)$ results to $O(N)$ for $N\geq 3$. As discussed above, the $O(N)$-invariant operators are given in \eqref{eq:4pt}, with Wilson coefficients $a_1$ and $a_2$, respectively. We then identify the independent operators that preserve $O(N-1)$ but break $O(N)$. These can be chosen to be proportional to
$X_{NN}^2$, $X_{NJ}X_N^J$, and $X_{NN}X^J_J$, where $J$ is summed over the remaining $N-1$ indices $J=1,\ldots,N-1$.
\begin{figure}[h!]
    \centering
    \includegraphics[width=0.8\linewidth]{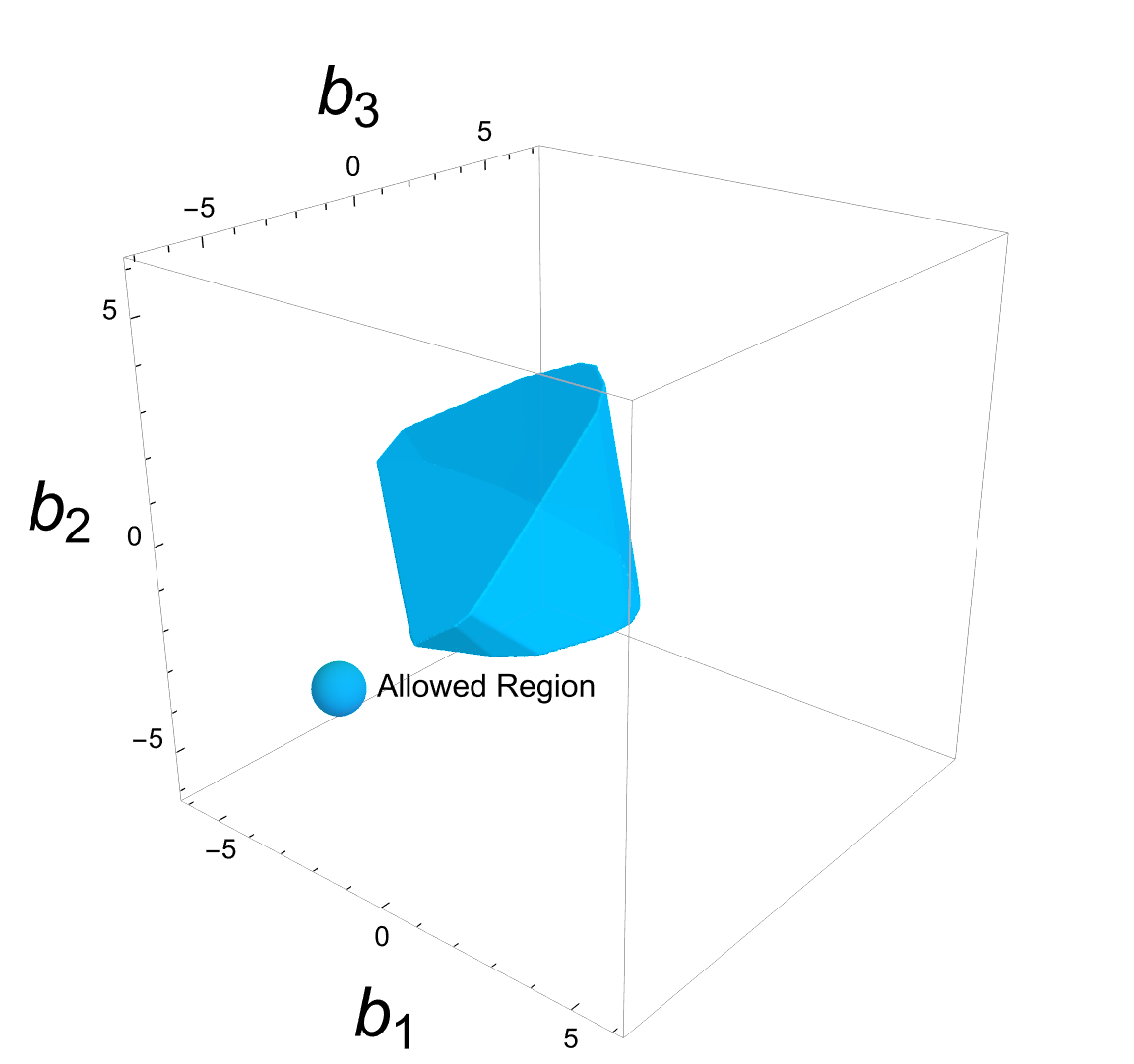}
    \caption{Allowed region (derived numerically) for the example given in \eqref{eq:O(N)toO(N-1)}, with $N=4$ fields and $O(4)$ symmetry preserving operators are governed by Wilsonian coefficients $a_1=2a_2=1$. The Wilson coefficients of the $O(3)$ operators are compact and bounded on all sides by the coefficients of the $O(4)$ operators. The more symmetry breaking directions are switched on and the more stringent the bounds.}
    \label{fig:ONON-1}
\end{figure}
We must first project out the $O(N)$--invariant components of these operators. The details of this calculation are given in Appendix~\ref{App:GAP}. We therefore introduce the following operators, which are $O(N-1)$--invariant and have no $O(N)$--invariant\footnote{Note that in the two-field example we had made a different basis choice, yet in both cases they obey the condition that the projections of the symmetry breaking operators onto $O(N)$ must vanish.} component:
\begin{equation}
\begin{aligned}
\O_1^{\ncancel{O(N)},\ O(N-1)}&=-\frac{1}{\Lambda^4}\( X_{NN}^2- \frac{1}{N(N+2)} \([X\nN{_N}]^2+2[X\nN{_N}^2]\)\)\,,\\
\O_2^{\ncancel{O(N)},\ O(N-1)}&=\frac{1}{\Lambda^4}\(X_{NJ}X_{N}^J-\frac{1}{(N+2)}\([X\nN{_N}^2]-\frac{1}{N}
[X\nN{_N}]^2\)\)\,,\\
\O_3^{\ncancel{O(N)},\ O(N-1)}&=\frac{1}{\Lambda^4}\(X_{NN}X^J_J-\frac{1}{N(N+2)}\((N+1)[X\nN{_N}]^2-2[X\nN{_N}^2]\)\)\,.
\end{aligned}
\end{equation}
Their Wilson coefficients are denoted by $b_{1,2,3}$ respectively.  We have introduced a minus sign in the definition of $\O_1^{\ncancel{O(N)},O(N-1)}$ for convenience. The EFT at this order, including all five operators, is then
\ba
\mathcal{L}_{\rm int}(\Phi)=\sum_{i=1}^2 a_i\O_i^{O(N)}+\sum_{i=1}^3 b_i\O_i^{\ncancel{O(N)},O(N-1)} \,.
\label{eq:O(N)toO(N-1)}
\ea
As an example, we take $N=4$ and fix the Wilson coefficients $a_1$ and $a_2$ of the $O(N)$--invariant operators. We then numerically determine the allowed region in the parameter space of the symmetry--breaking coefficients $b_i$. As shown in Fig.~\ref{fig:ONON-1}, this region is bounded in all directions.

\begin{figure}[h!]
    \centering
    \includegraphics[width=0.6\linewidth]{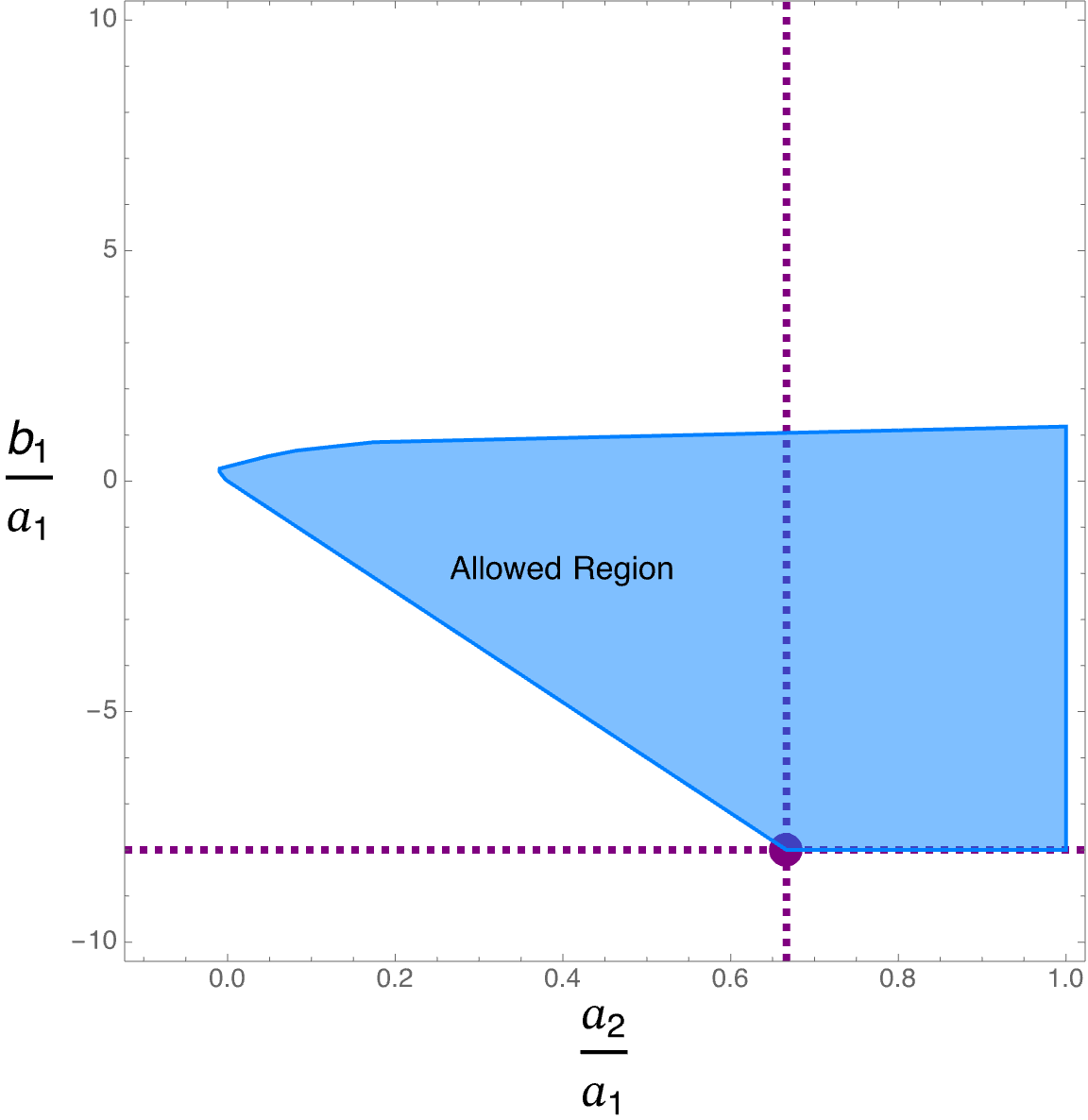}
    \caption{Slice of the parameter space for the $N=4$ theory with $b_2=b_3=0$. The dotted purple lines corresponds to $a_2=2a_1/3$ and $b_1=-8a_1$ respectively. Their intersection corresponds to the special case considered in \eqref{eq:LintN} where the field $\phi_N$ is fully decoupled and only one $\phi_N$ self-interaction is included. The special case shows the decoupled model as being precisely at the edge of the allowed region of parameter space. }
    \label{fig:Decoupled_general}
\end{figure}

\paragraph{Decoupled Sectors:}
Analogously to the two-field case, we could consider the case here where one of the fields, say field $\phi_N$, decouples. As an example we could consider the simple case
\begin{equation}
\label{eq:LintN}
\L_{\rm int}(\Phi)=\frac{\lambda_N}{\Lambda^4}(\partial \phi_N)^4
\end{equation}
which in our operator basis corresponds to setting $b_2=b_3=0$ and 
\begin{equation}
a_1=\frac{3}{N(N+2)} \lambda_N\,, \quad  a_2=\frac{2}{3}a_1\,, \quad{\rm and }\quad b_1=-\frac{N(N+2)}{3}a_1=-\lambda_N\,.
\end{equation}
We show a slice of the $b_2=b_3=0$  parameter space for $N=4$ in Fig.~\ref{fig:Decoupled_general}.
The special case where we only consider a single self-interaction \eqref{eq:LintN} is clearly covered by this analysis and lies at the edge of the allowed parameter space. 

\subsection{$O(8)\to U(4)\to \mathbb{Z}_2^4$}

We next consider the symmetry hierarchy
\ba
O(8)\to U(4)\to \mathbb{Z}_2^4\,.
\ea
This choice is motivated by the special role of $Sp(4)$ in supersymmetric Yang--Mills theories. Although $Sp(4)$ is the symmetry of primary interest in that context, the operators considered here do not distinguish its invariants from those of $U(4)$. Since we restrict our analysis to global symmetries, we therefore study the equivalent hierarchy $O(8)\to U(4)\to\mathbb{Z}_2^4$.\\

We begin with a set of $N=8$ real fields. The maximal level of global internal field space symmetry the EFT can enjoy is therefore $O(8)$ as given in \eqref{eq:4pt} (for $N=8$). A general EFT may include operators that preserve progressively  smaller subgroups of $O(8)$, down to operators with no non--trivial internal symmetry. We classify them according to their residual symmetry (if any). 

To identify the operators preserving $U(4)$, we introduce the antisymmetric tensor
\begin{equation}
    \Omega_{ab} = -\Omega_{ba}\,,
\end{equation}
which we can write in the following representation,
\begin{equation}
\Omega =
\begin{pmatrix}
0_{4\times4} & I_{4\times4}\\
-I_{4\times4} & 0_{4\times4}
\end{pmatrix}\,.
\end{equation}
The subgroup of $O(8)$ that preserves $\Omega$ is $U(4)$, and therefore contractions involving $\Omega$ can be used to construct $U(4)$--invariant operators. The resulting operator may also contain an $O(8)$--invariant component, which we remove using the group-averaging projector. This gives the genuinely $O(8)$--breaking, $U(4)$--preserving operator
\begin{equation}
\O^{\ncancel{O(8)},\ U(4)}=\frac{1}{\Lambda^4} \left(-\Omega_{ab} \Omega_{cd} X^{ac}X^{bd} - \frac{1}{7}\left( [X^2]-[X]^2\right)\right)\,.
\label{eq:U(4)}
\end{equation}
We note that the $U(4)$ operator breaks three of the eight independent $\mathbb{Z}_2$ symmetries, leaving five of them. The minus sign is for convenience. We now subtract any $O(8)$ invariant parts of this operator using the projection method discussed earlier. We know that $a_{1,2}>0$ (independently of the presence of any other operator), and we find that the exact constraint on this operator is two sided, and given by 
\begin{equation}
-\frac{7}{6}a_2<b_1<7a_2\,.
\label{U(4)bound}
\end{equation}
Note that these bounds are valid even when any other operators (in our operator basis) are switched on. 
We then add an operator that is invariant under the subgroup  $\mathbb{Z}_2^4$, but breaks both $O(8)$ and $U(4)$,
\begin{equation}
    \mathcal{O}^{\ncancel{O(8)},\ \ncancel{U(4)},\ \mathbb{Z}^4_2}=\frac{1}{\Lambda^4} X_{44}X_{45}\,.
\label{eq:O(8)d1}
\end{equation}
This operator contains no residual components invariant under either $U(4)$ or $O(8)$, so no further projection is needed.
The resulting sequence of operators associated with the symmetry hierarchy $O(8)\to U(4)\to \mathbb{Z}_2^4$ is therefore given by\footnote{We note that the exact residual symmetry of \eqref{eq:O(8)full} with all operators included is technically larger than $\mathbb{Z}_2^4$ -- it has an accidental extra $U(2) \times \mathbb{Z}_2$ symmetry.}. 
\begin{equation}
\L^{N=8}_{\mathbb{Z}_2^4}=\sum_{i=1}^2 a_i\, \O_{i}^{O(8)}+ b_1 \O^{\ncancel{O(8)},\ U(4)} + c_1 \O^{\ncancel{O(8)},\ \ncancel{U(4)},\ \mathbb{Z}_2^4} \,.
\label{eq:O(8)full}
\end{equation}
If we were to set $b_1=0$, then the Wilsonian coefficient $c_1$ is already strongly constrained by the coefficients of the maximally symmetric operators,
\begin{equation}
    |c_1|^2<8a_1a_2\,.
\end{equation}
When $b_1\ne 0$, this bound gets slightly modified as derived in Appendix~\ref{App:O8Bound}, with nominally $a_2\to a_2-b_1/7$. For the $U(4)$-breaking operator to be present, at least one operator that preserves a larger symmetry group must be present. More generally, in the absence of $c_1$, the bound on $b_1$ is given in \eqref{U(4)bound}, for which $a_1$ plays no role, so we $c_1$ is more strongly constrained than $b_1$. Note however that we still see that in the total absence of $O(8)$ operators (\ie when $a_2=a_1=0$), there can be no $O(8)$--breaking operator. We can find the bounds on $c_1$ analytically in term of $a_1,a_2,b_1$ for $b_1>0$ and $b_1<0$. The bounds and an outline of the derivation of these is given in Appendix~\ref{App:O8Bound}.

\begin{figure}[t] 
\centering 
\begin{subfigure}[t]{0.48\textwidth} 
\centering 
\includegraphics[width=\linewidth]{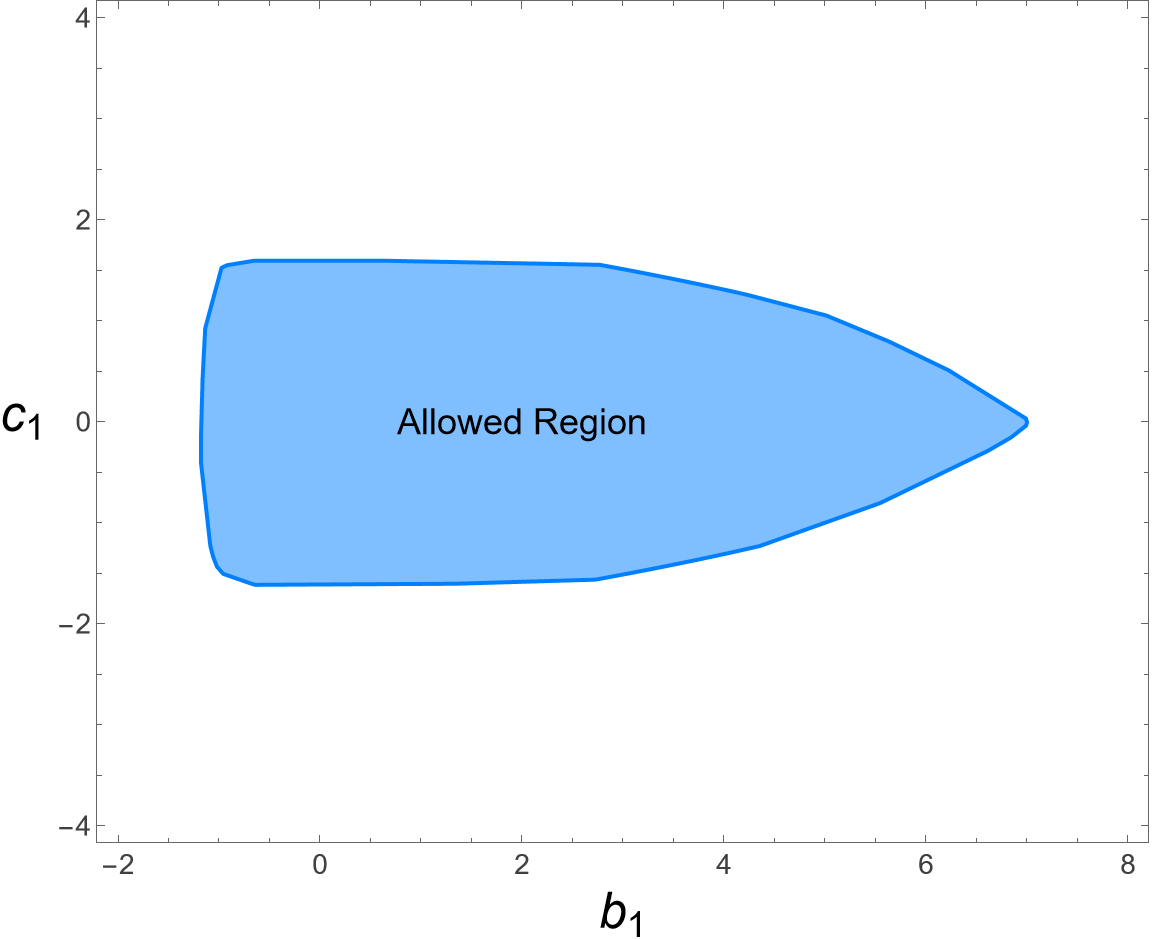} 
\caption{Allowed region for the 8-field EFT given in \eqref{eq:O(8)full} with symmetry hierarchy $O(8)\to U(4)\to \mathbb{Z}_2^4$.  Bounds are derived using $2-2$ amplitudes. The coefficients of the maximally symmetric $O(8)$ preserving operators are set to $a_2=2a_1=1$. $b_1$ parameterises the $U(4)$ preserving operator and $c_1$ the $ \mathbb{Z}_2^4$ invariant one. } 
\label{fig:O8full} 
\end{subfigure} 
\hfill 
\begin{subfigure}[t]{0.48\textwidth} 
\centering 
\includegraphics[width=\linewidth]{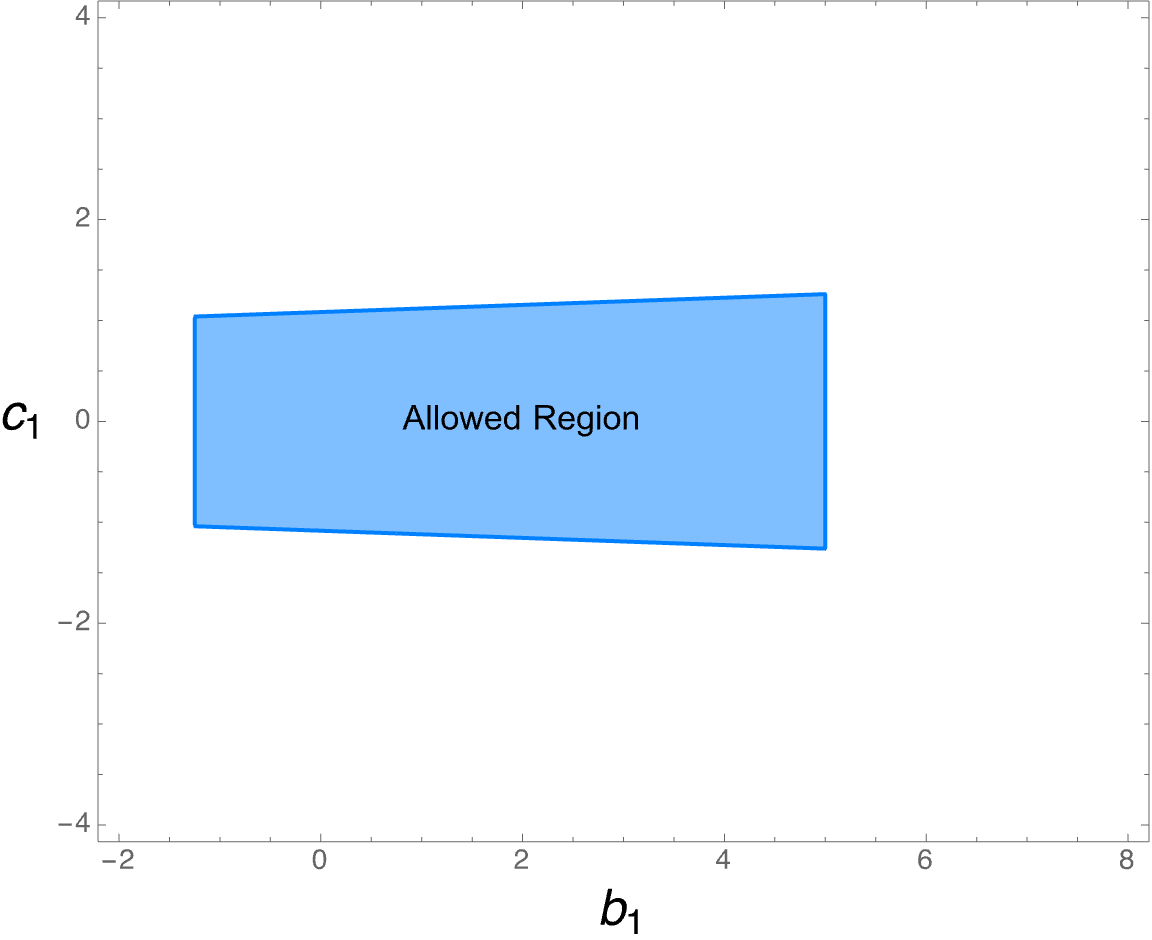} 
\caption{Allowed region for 7-field EFT given in \eqref{eq:O(7)full}, with $O(7) \to G_2 \to \mathbb{Z}_2^6$ symmetry hierarchy. Bounds are derived using $3-3$ amplitudes. The coefficients of the maximally symmetric $O(7)$ operators are set to  $a_2=a_3=2a_1=1$.  $b_1$ parameterises the exceptional Lie group $G_2$ preserving operator and $c_1$ the residual $\mathbb{Z}_2^6$ invariant one. 
} 
\label{fig:O7dim12symbreaking} 
\end{subfigure} 
\caption{Comparison of the allowed parameter space for EFTs exhibiting different symmetry-breaking hierarchies.} 
\label{fig:symmetrybreakingcomparison} 
\end{figure}

We can check that these are the optimal (linear) bounds, \ie they cannot be restricted any further by other choices on indefinite amplitude. We plot the resulting symmetry--breaking Wilson-coefficient parameter space in Fig.~\ref{fig:O8full}.

\subsection{$O(7) \to G_2 \to \mathbb{Z}_2^6$}

We next consider the case of an exceptional symmetry group. 
For the derivative interactions studied here, the first non-trivial compact exceptional Lie group arises at cubic order in $X^{ab}$, corresponding to six-derivative operators, for $N=7$ real scalar fields. In particular, we can construct a $G_2$-invariant operator that breaks the $O(7)$ symmetry. The $O(7)$-invariant operators are given in \eqref{eq:6pt}, with $N=7$, and positivity requires
\ba
a_1>0\,, \quad a_2>0\,, \quad a_3>0\,,
\ea
even in the presence of any additional operators in the EFT.

To construct the $G_2$--invariant, $O(7)$--breaking operator, we use the $G_2$-invariant three-form $\Phi_{abc}$ and its dual four-form $\psi_{abcd}$. We begin with the operator
\begin{equation}
    \O^{G_2}_{\text{dim}-12}=\frac{1} {\Lambda^8} \psi_{abcd}\psi_{fghd}X^{af}X^{bg}X^{ch}\,,
\label{eq:G2}
\end{equation}
then using the contraction identity given in \cite{Dwivedi_2025} and  separating out the contributions that are invariant under $O(7)$, the genuinely $O(7)$-breaking part can be expressed in terms of the $G_2$-invariant three-form as 
\begin{equation}
    \O^{G_2}_{\text{dim}-12}=\frac{1} {\Lambda^8} \Phi_{abc}\Phi_{def}X^{ad}X^{be}X^{cf}\,.
\end{equation}
This operator still contains an $O(7)$-invariant component. Subtracting that component using the group-averaging projector gives the genuinely $O(7)$-breaking, $G_2$-preserving operator \eqref{eq:G2app}
\begin{equation}
    \O^{\ncancel{O(7)},\ G_2}_{\text{dim}-12}=\frac{1}{\Lambda^8}\left( \Phi_{abc}\Phi_{def}X^{ad}X^{be}X^{cf}- \frac{1}{5}\left( [X]^3-3[X][X^2]+2[X^3]\right) \right)\,,
\end{equation}
with Wilson coefficient $b_1$. This gives exact bounds on $b_1$ 
\begin{equation}
    -\frac{5}{4}a_3<b_1<5a_3\,.
\label{eq:b1O7}
\end{equation}
Note that these bounds are valid even when any other operators (in our operator basis) are switched on. 
We can now add an EFT term which breaks both $O(7)$ and $G_2$ symmetries, given by 
\begin{equation}
    \O^{\ncancel{O(7)},\ \ncancel{G_2},\ \mathbb{Z}_2^6}=\frac{1}{\Lambda^8}X_{44}^2 X_{45}\,,
\end{equation}
so that we can consider the following EFT\footnote{We note that the exact residual symmetry of \eqref{eq:O(7)full} in fact is $\frac{SU(2) \times \mathbb{Z}_4}{\mathbb{Z}_2} \times \mathbb{Z}_2$ rather than just $\mathbb{Z}_2^6$.}
\begin{equation}
\L^{N=7}_{\mathbb{Z}_2^6}=\sum_{i=1}^2 a_i \O_{i}^{O(7)}+ b_1 \O^{\ncancel{O(7)},\ G_2} + c_1 \O^{\ncancel{O(7)},\ \ncancel{G_2},\ \mathbb{Z}_2^6} \,.
\label{eq:O(7)full}
\end{equation}
Using the full set of positivity bounds, we determine the allowed region for the symmetry-breaking Wilson coefficients, shown in Fig.~\ref{fig:O7dim12symbreaking}. The Wilson coefficient $b_1$ of the operator preserving the exceptional compact symmetry group $G_2$ is bounded within a compact region. The coefficient $c_1$ associated with the further breaking to the smaller subgroup $\mathbb{Z}_2^6$ is also compact and more severely constrained, independently of the value of $b_1$. Thus, this symmetry--breaking pattern exhibits the same hierarchical behaviour observed for the other subgroups of $O(N)$. In particular, preserving the exceptional compact Lie group $G_2$ does not enlarge the allowed parameter space for the operators that break $O(N)$, relative to the corresponding patterns involving the breaking of $O(N)$ to say the subgroup $O(N-1)$.

\section{Discrete Internal--Symmetry Hierarchies}
\label{sec:Discrete}
Previously, we studied symmetry hierarchies in which the first subgroup of the maximal $O(N)$ symmetry was continuous. We now consider hierarchies in which the $O(N)$ symmetry is reduced directly to a discrete subgroup, followed by a further reduction to a smaller discrete subgroup.

Every $O(N)$-invariant EFT considered here possesses $N$ guaranteed $\mathbb{Z}_2$ symmetries, corresponding to independent sign flips of the scalar fields. In this section, we will focus on operators that enter the $3-3$ amplitude at tree level. We identify operators that enjoy an initial $O(2)$ symmetry and consider operators that preserve different subgroups of these discrete symmetries. We compare the bounds on Wilson coefficients associated with operators that break different numbers of the $\mathbb{Z}_2$ symmetries. We find the same qualitative pattern as in the hierarchies involving continuous subgroups: operators preserving smaller discrete subgroups, and hence breaking more of the original $\mathbb{Z}_2$ symmetries, are more tightly constrained than those preserving larger discrete subgroups.
\begin{figure}[h!]
    \centering
    \includegraphics[width=0.9\linewidth]{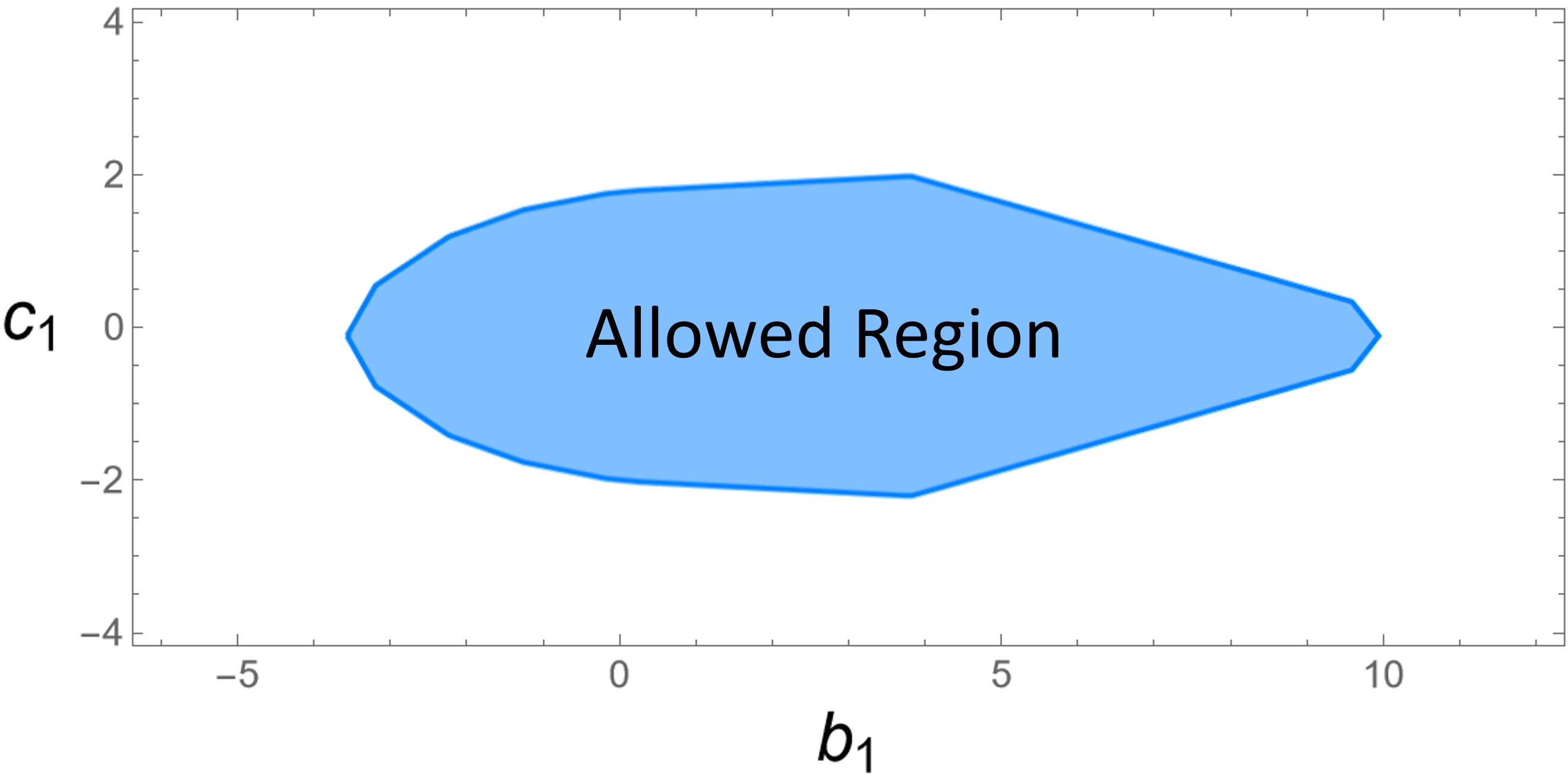}
    \caption{Bounds on the symmetry breaking parameter space of \eqref{eq:O(2)dim12} illustrating the hierarchy of discrete internal symmetries. As in the continuous--symmetry examples, the coefficient $c_1$, associated with the operator preserving the smaller discrete subgroup (here $\mathbb{Z}_2$), is more tightly constrained than $b_1$, which preserves the larger discrete subgroup (here $\mathbb{Z}_2\times \mathbb{Z}_2$). We have set the coefficients of the maximally symmetric $O(2)$ operators to $a_2=2a_1=1$.}
    \label{fig:dim12O(2)}
\end{figure}
We start with the six-derivative EFT given in \eqref{eq:6pt} with $N=2$. Note that in order to apply the positivity bounds we shall for now assume that no dimension-8 operators are present. For two fields, we have only two independent operators, $a_1,a_2>0$ that preserve the $O(2)$ symmetry. We include two more operators, the first of which is
\begin{equation}
    \mathcal{O}^{\ncancel{O(2)},\ \mathbb{Z}_2 \times \mathbb{Z}_2 }=\frac{1}{\Lambda^8}[X_{11}^2X_{22}-P_{O(2)}X_{11}^2X_{22}]\,,
\end{equation}
which breaks the global $O(2)$ symmetry, but enjoys a discrete $\mathbb{Z}_2 \times \mathbb{Z}_2$ symmetry. We also include
\begin{equation} 
\mathcal{O}^{\ncancel{O(2)},\ \ncancel{\mathbb{Z}_2 \times\mathbb{Z}_2},\ \mathbb{Z}_2 }=\frac{1}{\Lambda^8}X_{11}^2X_{12}\,,
\end{equation}
which breaks the $O(2)$ symmetry and enjoys only a single $\mathbb{Z}_2$ symmetry. This operator already contains no $O(2)$ invariant pieces. The full EFT is 
\begin{equation}
\L^{N=2}_{\mathbb{Z}_2, \text{dim-12}}=\sum_{i=1}^2 a_i \O_{i}^{O(2)}+ b_1 \O^{\ncancel{O(2)},\ \mathbb{Z}_2 \times\mathbb{Z}_2} + c_1 \O^{\ncancel{O(2)},\ \ncancel{\mathbb{Z}_2 \times\mathbb{Z}_2},\ \mathbb{Z}_2} \,.
\label{eq:O(2)dim12}
\end{equation}
The symmetry breaking parameter space is shown in Fig.~\ref{fig:dim12O(2)}.
We can clearly see a similar pattern when the starting global symmetry is broken directly to discrete subgroups. This example focused on $N=2$ for concreteness but the same pattern  holds for general $N$ without loss of generality. 

\section{Conclusions}
\label{sec:Conclusion}

When the number of free couplings in an EFT becomes very large, it can quickly become difficult to constrain the allowed region properly using physical considerations. Understanding the constraints on symmetry-breaking Wilson coefficients imposed by positivity bounds is therefore a well-motivated direction of study. In this work, we have investigated the interplay between internal symmetry breaking and positivity constraints in scalar EFTs. Our primary goal was to determine whether positivity arguments can provide non-trivial bounds on coefficients associated with explicit symmetry breaking. A key aspect of our approach was the use of indefinite external states, which probe linear combinations of scattering channels, providing access to constraints that are not immediately visible otherwise. \\

We have demonstrated a clear pattern of the restrictions on operators which break an internal symmetry group. We built our operator basis based on the level of symmetry they preserve and   found that all coefficients of operators that break the internal symmetry  are constrained by those that preserve it.
Additionally, the strength of the bounds on such terms is directly correlated with the number of internal symmetries that are broken. We demonstrated the consequences of the explicit breaking of successively smaller internal subgroups of the original symmetry on the allowed coefficient parameter space. We showed this extensively using several symmetry breaking sequences,  starting with $O(N)\to O(N-1)$ and progressively moving onto other symmetry hierarchies including for instance $O(8) \to U(4)  \to \mathbb{Z}_2^4$. We also consider groups that contain exceptional subgroups as in $O(7) \to G_2 \to \mathbb{Z}_2^6$ and also investigated hierarchies between discrete subgroups. In all cases, the smaller the size of the subgroup, the stronger the bound on the associated coefficient.\\

This general pattern has the potential to be a powerful insight to investigate EFTs  based on symmetry considerations alone. Our results indicate that positivity bounds can be viewed not only as constraints on EFT coefficients, but also as probes of the organisation of internal symmetries themselves. The hierarchy of residual symmetries plays a direct role in determining the allowed parameter space. This perspective may prove particularly useful in theories with large global symmetries, exceptional groups, or multiple stages of symmetry breaking, where conventional EFT analyses can become cumbersome. 
More importantly, the patterns we find suggest that these bounds may be pointing towards a deeper principles where consistency of the UV favors  enhanced--symmetry theories, or at least restricts how far a consistent low-energy theory can depart from them, even when the enhanced symmetry was not apparent in the premises used to construct the internal field space.\\

While this analysis focused on special patterns for concreteness, our results can be generalised in multiple directions which is left for future work. One natural extension is to incorporate particles with spin. Another is to consider gauge symmetries. It would also be interesting to explore whether the relations identified here admit a more general group-theoretic interpretation. Another interesting extension would be to understand the large $N$ limit on the positivity bounds for a variety of EFTs with $N$ fields. In our case of starting with $O(N)$ symmetry, we can include operators which break $O(N)$ but preserve $\mathbb{Z}_2$ in our basis. One way to do this is to include all possible operators of the form $X_{aa}X_{ab}$ with arbitrary coefficients, given schematically by 
\begin{equation}
\mathcal{L}_{\text{dim}-8}^{\ncancel{O(N)},\ \mathbb{Z}_2}= \sum_{a \neq b}b_{ab} X^{aa}X^{ab} \,,
\end{equation}
and all of the projectors of these are zero since the results of the integrals are proportional to $\delta_{ab}$ but we sum over $a \neq b$. However the proper normalisation of the state will scale as $\sqrt{N-1}$, which one would then expect to enter the bound for the associated Wilsonian coefficient.   We leave the rigorous classification of these operators in the large $N$ limit and the implied bounds for further work. Additionally, using null constraints as given in \cite{Du:2021byy} or applying different classes of bounds as given in \cite{Li:2021lpe} are also interesting avenues for further work.   \\

Overall, this work provides evidence that positivity constraints offer a powerful framework for studying explicitly the internal symmetry structure of  EFTs. By using indefinite external states, and the systematic classification of residual symmetries, we obtain a picture in which consistency conditions impose non-trivial restrictions on the magnitude of the coefficients of symmetry-breaking operators. We expect these ideas to have applications in a wide range of EFTs and to provide further insight into the connection between symmetry and ultraviolet completeness.

\paragraph{Acknowledgements:} We thank Andrew Tolley for great discussions. AB is supported by an
STFC studentship. The work of CdR is supported by STFC Consolidated Grant ST/X000575/1 and by STFC Consolidated Award UKRI4230: Abdus Salam Centre for Theoretical Physics at Imperial College London. CdR is also supported by a Simons Investigator award SFI-MPS-SIP-00017355.

\appendix

\section{Dispersion Relations and Superposition Positivity Bounds for $2-2$ scattering}
\label{App:2-2}

We review here the standard derivation of positivity bounds using the dispersion relations for $2-2$ elastic scattering amplitudes.  We assume the particle spectrum is composed of massless scalars and consider scattering between indefinite states which are defined as 
\begin{equation}
\label{eq:Psi}
\begin{aligned}
\left|\Psi\right\rangle&=\sum_{a=1}^N\alpha_a\left|\phi_a\right\rangle\,,\\
{\rm with} \qquad \alpha_a&=\(\prod_{b<a}e^{i \varphi_b}\sin \theta_b\) \times
\begin{cases}
\cos \theta_a\ & \ a<N\\
1 \ & \ a=N\,,
\end{cases}
\end{aligned}
\end{equation}
where the $\left|\phi_a\right\rangle$ denotes the definite field state $a$ (with implicit momentum). \\

We start with the assumptions that a standard Wilsonian UV completion is local, obeys Lorentz invariance, causality and unitarity. Applying these culminates in the dispersion relation which expresses the full $2-2$  scattering amplitude in terms of its discontinuities. This relies on the definition of the physical scattering amplitude as a limit of the complexified amplitude, which is a function of complex Mandelstam variables. One can show that this amplitude is analytic modulo poles and specific branch cuts on the real axis only. Using Cauchy's integral formula, we can write the amplitude as a contour integral over the usual deformed contour, giving
\begin{equation}
\mathcal{A}(s, t) = \frac{1}{2\pi i} \oint_{\mathcal{C}} \frac{A(s_0, t)}{(s_0 - s)} \, \d s_0 + \text{poles}\,,
\end{equation}
and considering the pole--subtracted amplitude, it can be shown, using the Froissart-Martin bound \cite{Froissart:1961ux,Jin:1964zza,Martin:1965jj,Martin:1969ina} that 
\begin{equation}
\begin{gathered}
\partial ^2_s\mathcal{A}(s,t)=\frac{2}{\pi} \int_{4m^2}^{\infty} \d s_0\ \left\{\frac{\text{Disc}_{s_0}\mathcal{A}_s(s_0,t)}{(s_0-s)^3} + \frac{\text{Disc}_{s_0}\mathcal{A}_u(s_0,t)}{(s_0-u)^3}\right\} + \partial^2_s\mathcal{I}(s,t)\,,\\
\text{where}\quad
\partial^2_s\mathcal{I}(s,t) = 2\int_{|s_0| \to  \infty}\d s_0\frac{\mathcal{A}(s_0,t)}{(s_0-s)^3}\to 0\,.\qquad \qquad
\end{gathered}
\end{equation}
Unitarity implied positivity of the discontinuity in the physical region, \ie on the right hand cut, which by crossing symmetry can then be related to positivity of the discontinuity of the left hand cut, yielding the familiar positivity bound 
\begin{equation}
\partial_s^2\mathcal{A}(s,t) > 0  \quad 0\le s \le 4m^2\,.
\end{equation}
In principle the bounds can be extended away from the forward limit \cite{deRham:2017avq}, though in this work we shall only make use of forward positivity bounds.  A version of this bound holds for superposition states even for fields of different masses \cite{deRham:2025htd}. 

\section{Group Averaging Projector}
\label{App:GAP}

In this appendix, we elaborate on the calculations involving the group averaging projector used in this paper. 
\subsection{$O(N)\to O(N-1)$}

For the operator $X_{NN}^2$, the projector is
\begin{equation}
    \mathcal{P}_{O(N)}X_{NN}^2=\int_{O(N)} \text{d}A [A_{Na}A_{Nb}A_{Nc}A_{Nd}X_{ab}X_{cd}]\,,
\end{equation}
and for orthogonal matrices, we use the identity 
\begin{equation}
    \int_{O(N)} \text{d}A\  A_{Na}A_{Nb}A_{Nc}A_{Nd}=\frac{
\delta_{ab}\delta_{cd}
+\delta_{ac}\delta_{bd}
+\delta_{ad}\delta_{bc}
}{N(N+2)}\,,
\end{equation}
which therefore implies
\begin{equation}
    \mathcal{P}_{O(N)}X_{NN}^2=\frac{1}{N(N+2)}\left( [X]^2+2[X^2]\right)\,.
\end{equation}
For the operator $X_{NJ}X_{N}^J$, the projector is 
\begin{equation}
    \mathcal{P}_{O(N)}X_{NJ}X_{N}^J=\sum_{i=1}^{N-1}\int_{O(N)} \text{d}A\, [A_{Na}A_{Jb}A_{Nc}A^J_d X^{ab}X^{cd}]\,,
\label{eq:XNJ2}
\end{equation}
which can be simplified using the orthogonality relation, 
\begin{equation}
    \sum_{i=1}^{N-1}A_{Jb}A^J_d=\delta_{bd}-A_{Nb}A_{Nd}\,.
\end{equation}
Then bearing in mind that
\begin{equation}
    \int_{O(N)} \text{d}A A_{Na}A_{Nb}=\frac 1N\, 
\delta_{ab}\,,
\end{equation}
we obtain 
\begin{equation}
    \mathcal{P}_{O(N)}X_{NJ}X_{N}^J=\frac{1}{N(N+2)}\left(N[X^2]-[X]^2\right)\,.
\end{equation}
Finally, the projector of $X_{NN}X^J_J$ is written as 
\begin{equation}
    \mathcal{P}_{O(N)}X_{NN}X^J_J=\sum_{i=1}^{N-1}\int_{O(N)} \text{d}A [A_{Na}A_{Nb}A_{Jc}A^J_dX^{ab}X^{cd}]\,,
\end{equation}
and using orthogonality as before, we obtain 
\begin{equation}
     \mathcal{P}_{O(N)}X_{NN}X^J_J=\frac{1}{N(N+2)}\left((N+1)[X]^2-2[X^2]\right)\,.
\end{equation}
As an example, we present the exact analytic bounds on $b_1$ when the other $O(N-1)$ operators are turned off (when $b_2,b_3=0$), with $a_{1,2}>0$. We obtain the lower bound 
\begin{equation}
b_1>-N(N+2)\min \left\{\frac{a_1}{3}, \frac{a_2}{2}\right\}\,,
\end{equation}
and the the upper bound 
\begin{equation}
b_1< \min\left\{\frac{N(N+2)a_1}{N(N+2)-3},\, 
\frac{N(N+2)\, \mathcal{S}}{2\left[3N(N+2)-8\right]} 
 \right\}\,,
\end{equation}
with
\ba
\mathcal{S}= \sqrt{ \left(\left[N(N+2)-4\right]a_1-2a_2\right)^2 + 4\left[3N(N+2)-8\right]a_2(2a_1-a_2) }
- \left[N(N+2)-4\right]a_1+2a_2\,,\nn
\ea
and the numerical results are presented for $N=4$ in Fig.~\ref{fig:ONON-1}.

\subsection{$O(8) \to U(4)$}

We calculate the projector of $-\Omega_{ab}\Omega_{cd}X^{ac}X^{bd}$ to obtain the $U(4)$ operator given in \eqref{eq:U(4)}. Since 
\begin{equation}
\mathcal{P}_{O(8)}\Omega_{ab}\Omega_{cd}=\frac{1}{7}(\delta_{ac}\delta_{bd}-\delta_{ad}\delta_{bd})\,,
\end{equation}
and the result of the total projection must be a linear combination of the two $O(8)$ invariants $[X]^2,[X^2]$, contracting this identity with $X_{ac}X_{bd}$ gives the result 
\begin{equation}
\mathcal{P}_{O(8)}(-\Omega_{ab}\Omega_{cd}X^{ac}X^{bd})= \frac{1}{7}([X^2]-[X]^2)\,.
\end{equation}

\subsection{$O(7) \to G_2$}

We can show that the 7-dimensional projector operator can be expressed as
\begin{equation}
    \mathcal{P}_{O(7)}\Phi_{abc}\Phi^{def}\propto \delta^{abc}_{def} \,,
\end{equation}
where $\delta^{abc}_{def} $ is the generalised Kronecker delta. This is the identity on the space of three-forms under $O(7)$, and hence it is a totally antisymmetric tensor. Then we can define 
\begin{equation}
T_{abc,def}\equiv
\int_{O(7)}\d A\,
(A\cdot\phi)_{abc}(A\cdot\phi)_{def}\,.
\end{equation}
The invariance of the integration measure implies that $T$ commutes with the elements of $O(7)$. By Schur's lemma, $T$ must therefore be proportional to the identity on this space, whose kernel is represented by the totally antisymmetric generalised Kronecker delta. To find the constant of proportionality, we use the fact that 
\begin{equation}
    \Phi_{abc}\Phi^{abc}=42, \quad \delta^{abc}_{abc}=210\,,
\end{equation}
so contracting with $X_{ad}X_{be}X_{cf}$ gives the operator 
\begin{equation}
\label{eq:G2app}
    \O^{\ncancel{O(7)},\ G_2}_{\text{dim}-12}=\frac{1}{\Lambda^8}\left( \Phi_{abc}\Phi_{def}X^{ad}X^{be}X^{cf}- \frac{1}{5}\left( [X]^3-3[X][X^2]+2[X^3]\right) \right)\,.
\end{equation}

\section{Analytic Result for $O(8) \to U(4) \to \mathbb{Z}_2^4$}
\label{App:O8Bound}

Even in the presence of other operators, one can check that the bounds defined in \eqref{U(4)bound} remain valid \ie $(-7/6) a_2<b_1<7a_2$. So defining the shifted coefficient
\ba
\tilde a_2=a_2-\frac{b_1}{7}>0\,,
\ea
so the bound \eqref{U(4)bound} which can be written as 
\ba
\label{eq:tilde a2}
\tilde a_2> 0 \quad{\rm and }\quad b_1>-\tilde a_2
\ea
remains valid in what follows. 

Now to study the bounds on the operator $\frac{1}{\Lambda^4} X_{44}X_{45}$, with coefficient $c_1$, it will prove convenient to consider elastic the scatterings between the two following states
\begin{equation}
\begin{aligned}
|\Psi\rangle_A
&= \sin{\chi} \sin{(\theta_A- \nu)} |\phi_1 \rangle + \cos\chi \cos{\theta_A} |\phi_4 \rangle + \cos\chi \sin{\theta_A} |\phi_5 \rangle - \sin{\chi} \cos{(\theta_A- \nu)}|\phi_8 \rangle,\\
|\Psi\rangle_B
&= -\sin{\chi} \sin{(\theta_B+ \nu)} |\phi_1 \rangle + \cos\chi \cos{\theta_B} |\phi_4 \rangle + \cos\chi \sin{\theta_B} |\phi_5 \rangle + \sin{\chi} \cos{(\theta_B+ \nu)}|\phi_8 \rangle\,,
\end{aligned}
\end{equation}
and define $\theta_A=(x+\Delta)/2$, $\theta_B=(x-\Delta)/2$.
 We can then write the positivity bounds as 
\ba
F&=&\tilde a_2 + (2a_1-\tilde a_2)\(\cos^2\!\chi\, \cos{\Delta} - \sin^2\!\chi\, \cos{(\Delta-2\nu)}\)^2 \nn\\
&+& 4 b_1  \cos^2\!\chi\sin^2\!\chi\, \cos^2\!(\Delta-\nu) \nn\\
&+& \frac{c_1}{2} \cos^4\!\chi\, \sin x \(\cos \Delta+\cos x\) >0\,.
\ea

\paragraph{$\bullet$ For $b_1 \ge -\tilde a_2/4$,}  we choose $\chi=0$, and obtain
\begin{equation}
F=\tilde a_2+ (2a_1-\tilde a_2) \cos^2{\Delta} + \frac{c_1}{2} \sin{x} (\cos{\Delta}+\cos{x})>0,
\end{equation}
yielding 
\begin{equation}
|c_1|^2< \min_{\Delta, x}  2 \left( \frac{\tilde a_2+ (2a_1-\tilde a_2) \cos^2{\Delta}}{\sin{x} (\cos{\Delta}+\cos{x})}\right),
\end{equation}
which can be extremised as
\begin{equation}
|c_1|<\min \left( \frac{16 a_1}{3 \sqrt{3}}, 4 \tilde a_2 \sqrt{1-\frac{\tilde a_2}{2 a_1}}   \right)\,,
\end{equation}
with the intersection of the two candidate bounds at $a_1=\frac{3 \tilde a_2}{4}$.

\paragraph{$\bullet$ For $b_1 < -\tilde a_2/4$,} we first note that since $b_1>-\tilde a_2$ from \eqref{eq:tilde a2}, so in this region where $b_1=-|b_1|$ is negative, $|b_1|$ is bounded on both sides $\tilde a_2<|b_1|<\tilde a_2/4$.
The bounds depends on whether  $\tilde a_2/4<|b_1| < \tilde a_2/2$ or $\tilde a_2/2<|b_1| < \tilde a_2$. 
\ba
|c_1|^2<\begin{cases}
\frac{2^{12}}{3^3} \frac{|b_1|}{\tilde a^2_2} \(\tilde a_2 +b_1\)^3\( 1- \frac{\tilde a_2}{2a_1}\)\,,  & \frac{\tilde{a}_2}{4}<|b_1|\leq \frac{\tilde{a}_2}{2} 
\qquad \ \ \ {\rm and }\quad a_1> \dfrac{\tilde{a}_2(\tilde{a}_2+b_1)}{2(\tilde{a}_2+2b_1)}\\
\frac{2^{10}}{3^3}(\tilde a_2 +b_1)^2 \left( 1-\frac{\tilde a_2 +b_1}{2a_1}\right)^2\,,  & 
\begin{cases}
\frac{\tilde{a}_2}{4}<|b_1|\leq \frac{\tilde{a}_2}{2} 
& \quad {\rm and }\quad a_1< \dfrac{\tilde{a}_2(\tilde{a}_2+b_1)}{2(\tilde{a}_2+2b_1)}\\
 \frac{\tilde{a}_2}{2}<|b_1|\leq \tilde{a}_2 & \quad {\rm and }\quad \forall\   a_1>0\,.
\end{cases}
\end{cases}
\ea
We can plot the numerical and analytic results together, showing different slices of parameter space, and showing excellent agreement between both.
\begin{figure}[h] 
\centering 
\begin{subfigure}[t]{\textwidth}
\begin{subfigure}[t]{0.48\textwidth} 
\centering 
\includegraphics[width=\linewidth]{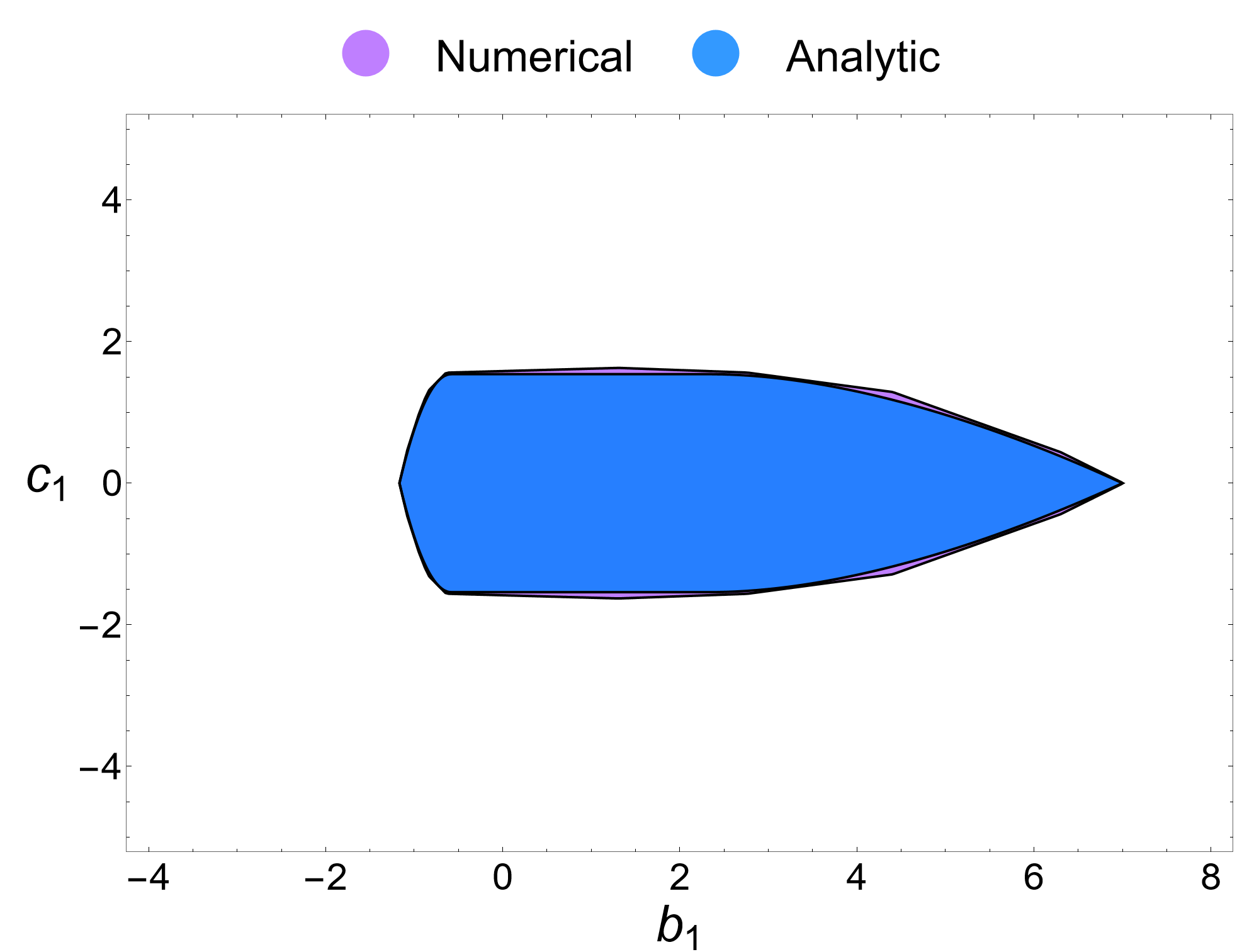} 
\end{subfigure} 
\hfill 
\begin{subfigure}[t]{0.48\textwidth} 
\centering 
\includegraphics[width=\linewidth]{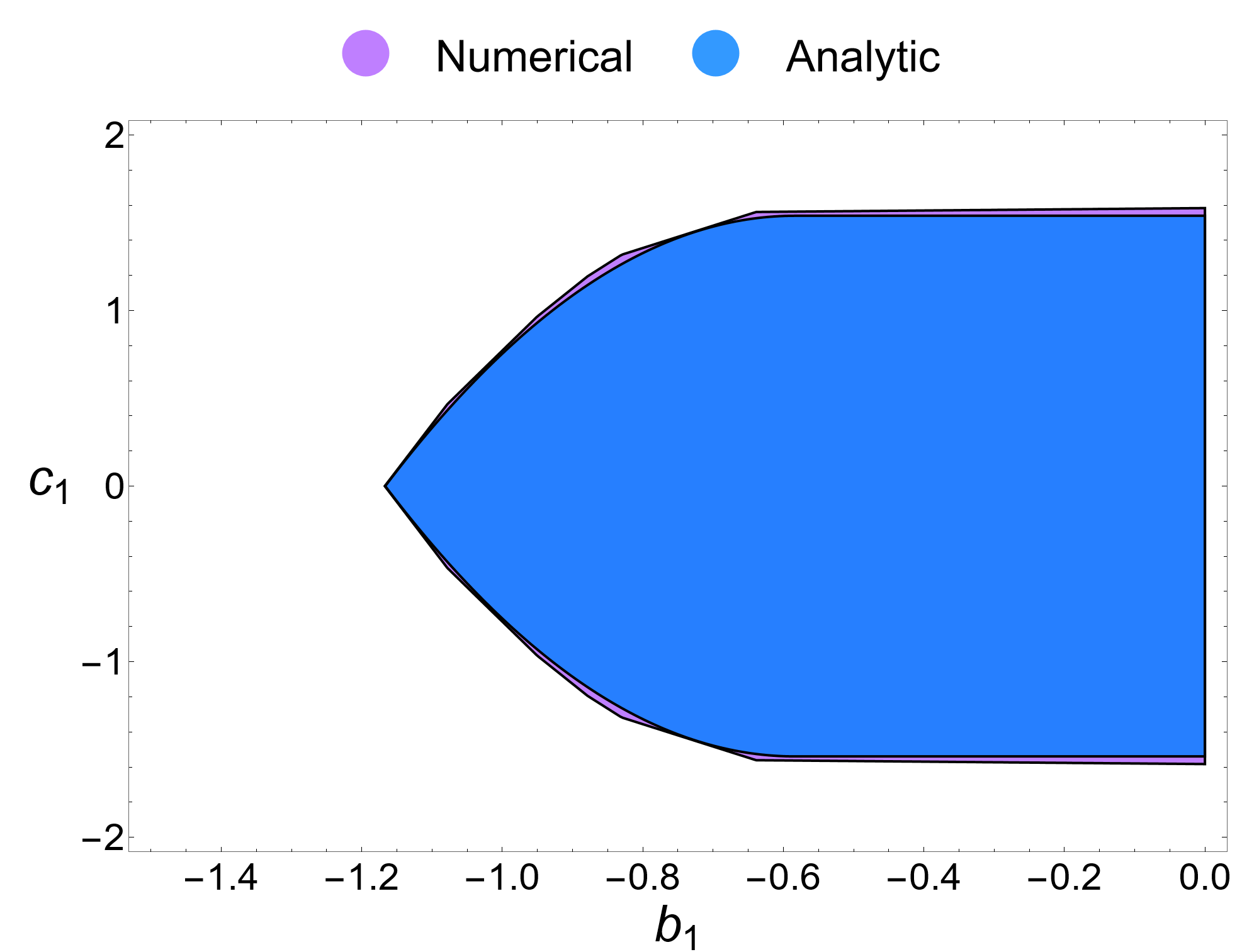} 
\end{subfigure} 
\caption{$a_2=2a_1=1$} 
\label{fig:plot_a1_12} 
\end{subfigure}

\vspace{0.5cm}

\begin{subfigure}[t]{\textwidth}
\begin{subfigure}[t]{0.48\textwidth} 
\centering 
\includegraphics[width=\linewidth]{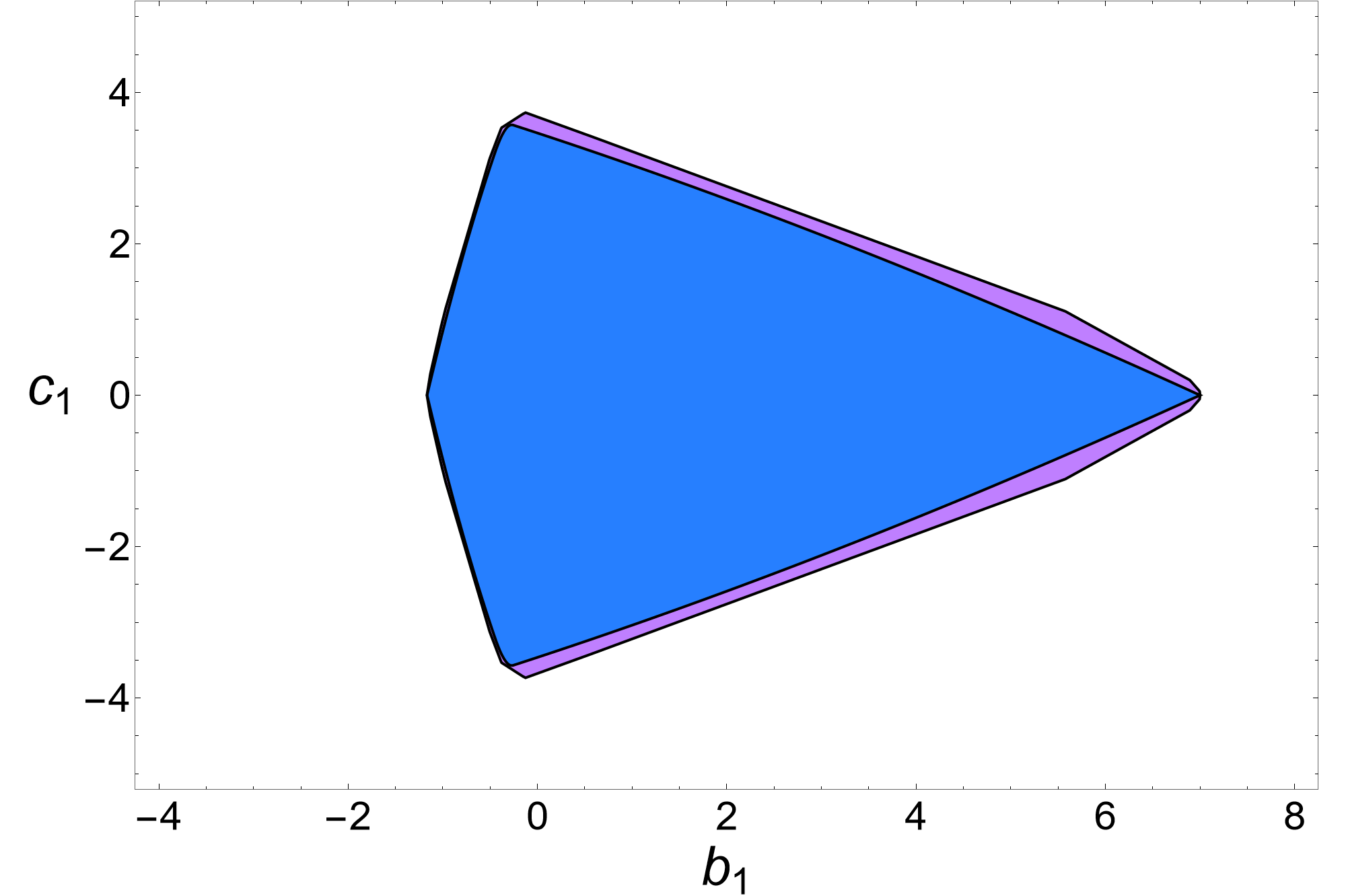} 
\end{subfigure} 
\hfill 
\begin{subfigure}[t]{0.48\textwidth} 
\centering 
\includegraphics[width=\linewidth]{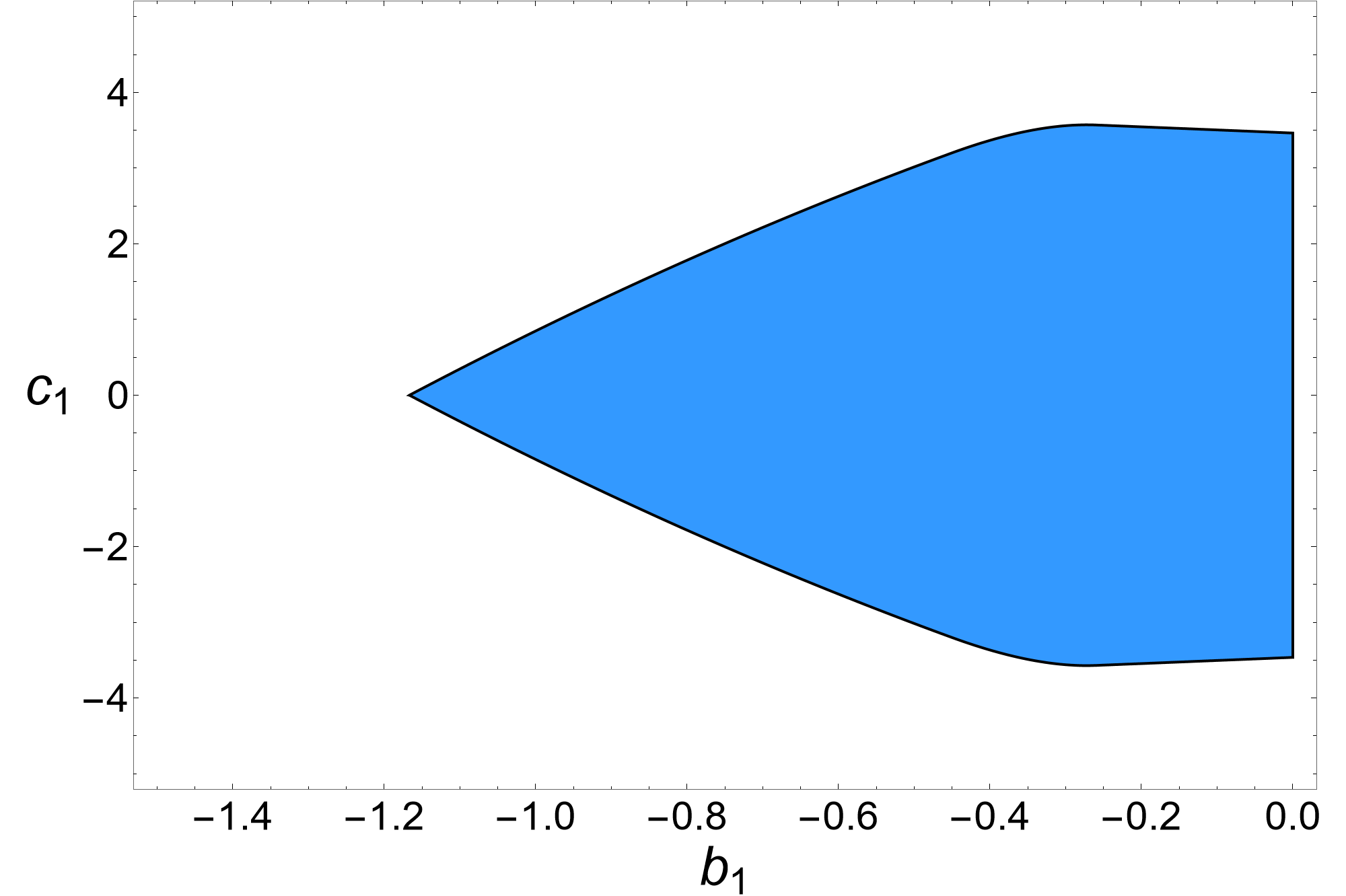} 
\end{subfigure} 
\caption{$a_1=2a_2=2$} 
\label{fig:plot_a1_2} 
\end{subfigure}
    \caption{Different slices of the constrained parameter space, evaluated numerically and analytically for the symmetry hierarchy $O(8) \to U(4) \to \mathbb{Z}_2^4$. The Wilsonian coefficients of the maximally symmetric $O(8)$ operators are given by the $a_i$. $b_1$ is the Wilsonian coefficient of the $U(4)$-preserving operator and $c_1$ that of the $\mathbb{Z}_2^4$ one. The plots on the right--hand side focus on the behaviour when $b_1<0$. We can see that the parameter space is still constrained to a single point at both the negative and positive maxima of $b_1$ and excellent agreement between the numerical constraints and the optimal analytic ones.}
    \label{fig:fourplots}
\end{figure}
We can see from Fig.~\ref{fig:fourplots} that our numerical results perform well against the known analytic results, and that the parameter space is reduced to a single point at either end of the allowed range for $b_1$.

\section{Crossing Symmetry for $3-3$ scattering}
\label{App:Crossing}

We will now outline the applications of the $n-n$ amplitude bound to superposition states, choosing the same kinematics and forward limit as \cite{Chandrasekaran:2018qmx}. We have $n(2n-3)$ independent Mandelstam variables 
\begin{equation}
    s_{ij}=-(p_i+p_j)^2=-2p_i \cdot p_j\,,
\end{equation}
and $s$ is chosen as 
\begin{equation}
    s=-(p_1+...+p_n)^2\,.
\end{equation}
We work in the limit where all fields can be treated as massless. In principle one could further generalise this analysis to include different mass eigenstates as performed in \cite{deRham:2025htd} though we expect corrections to be parametrically suppressed by the mass difference over the cutoff at least for tree-level completions. For more generic completions the situation is more delicate but procedures can be performed to properly account for  the failure of crossing symmetry. Even in the simplest case where all fields are treated as massless, the equivalent of the forward limit from $2-2$ to $3-3$ is non-trivial.  Considering the case of odd $n$, the chosen forward kinematics are
\begin{equation}
\begin{split}
p_{i} &= p_1 \quad \text{for $i=1$ mod $2$}, i \in [1,n-1]\\
p_{i} &= p_2 \quad \text{for $i=0$ mod $2$}, i \in [1,n-1]\\
p_i&=-p_{i+n}, i \in [1,n]\,.
\end{split}
\end{equation}
With these kinematics, for a single field $P(X)$ theory with the first non-negligible interaction $\lambda_nX^n$, the $n-n$ amplitude is given by 
\begin{equation}
    \mathcal{A}(s)=\left( \frac{2}{n-1}\right)^{n-3} n!(n-1)! \delta^{n-2}\lambda^n (s_n-\delta)^2\,,
\end{equation}
where the choice has been made that $s_{1n}=s_{2n}$ which we call $s_n$. Holding $\delta=(n-1)^2\frac{s_{12}}{4}>0$ constant, one can take an analytic continuation over $s_n$. The amplitude can then be written as 
\begin{equation}
\mathcal{A}(s)=\left( \frac{2}{n-1}\right)^{2n-3}n! (n-1)! \delta^{n-2}\lambda_n(s-\delta)^2\,.
\end{equation}
Performing the appropriate contour integrals given in \cite{Chandrasekaran:2018qmx}, and noting that in our convention, $X=-(\partial \phi)^2$, one finds that 
\begin{equation}
    \lambda_n=C(n) \int_{s_0}^\infty \frac{\text{d}s}{s^3}\left[\sum_Y \int \text{dLIPS}_Y|\mathcal{A}(\ket{n,s} \to Y)|^2+ \sum_Y\int\text{dLIPS}_Y|\mathcal{A}(\ket{n,s+2 \delta} \to Y)|^2 \right]>0\,,
\end{equation}
where $C(n)$ is a positive function of $n$, $\mathcal{A}$ is the amplitude from an $n$ particle state to a final state $Y$, and we are summing over all possible final states. We have used the Lorentz invariant phase space measure given by 
\begin{equation}
    d\mathrm{LIPS}_Y
=
(2\pi)^D
\delta^{(D)}\!\left(
\sum_{i=1}^{n} p_i + p_Y
\right)
\prod_{j\in Y}
\frac{\d^{D-1}{p}_j}
{(2\pi)^{D-1}\,2E_j}\,.
\end{equation}
We then utilise this result using an elastic amplitude between superposition states where all scalars are massless to infer the resulting positivity bounds. 
There is a subtle point between these steps, which is the use of crossing symmetry as defined with the kinematic choices. 

For $2-2$ scattering, one can show that crossing symmetry is manifest in the superposition amplitude by swapping $p_1$ and $p_3$ in each constituent amplitude and showing the total superposition amplitude is invariant. Considering a two single indefinite states containing two species, given by 
\begin{equation}
 \ket{\Psi}_{AB} = \ket{\Psi}_{A} \otimes   \ket{\Psi}_{B}\,,\quad {\rm with} \quad
\begin{cases}
    \ket{\Psi}_A=\sum_{a=1}^2 \alpha_a \ket{\phi_a} \\
    \ket{\Psi}_B=\sum_{a=1}^2 \beta_a \ket{\phi_a}\,,
\end{cases}
\end{equation}
with normalized indefinite coefficients as indicated in \eqref{eq:Psi}, $\sum|\alpha|^2=1=\sum|\beta|^2$.
The amplitude can then be written as a sum over constituent $2-2$ amplitudes 
\begin{equation}
    \mathcal{A}_{AB\to AB} = \sum_{i,j} \sum_{i', j'} \alpha_i \beta_j \alpha_{i'} \beta_{j'}\mathcal{A}_{ij \to i' j'}(p_1,p_2;p_3,p_4) \,,
\end{equation}
where $i,j$ label incoming species and $i',j'$ label outgoing species. We can then write 
\begin{equation}
 \mathcal{A}_{AB\to AB}^{\text{crossed}} = \sum_{i,j} \sum_{i', j'} \alpha_i \beta_j \alpha_{i'} \beta_{j'}\mathcal{A}_{ij' \to i' j}(-p_3,p_2;p_1,p_4) \,,
\end{equation}
and we are free to relabel the species indices since we are summing over all of them, giving 
\begin{equation}
 \mathcal{A}_{AB\to AB}^{\text{crossed}} = \sum_{i,j} \sum_{i', j'} \alpha_i \beta_j \alpha_{i'} \beta_{j'}\mathcal{A}_{ij \to i j}(-p_3,p_2;p_1,p_4) \,,
\end{equation}
and finally, applying the forward limit to both the amplitude and the amplitude after we apply crossing, we find that 
\begin{equation}
    \mathcal{A}_{AB\to AB}^{\text{crossed}} = \mathcal{A}_{AB\to AB}\,.
\end{equation}
Similarly, with the help of the $n-n$ forward limit, this can be extended to the $n-n$ superposition amplitude. We will sketch this out for $3-3$ scattering with superposition states. The amplitude is 
\begin{equation}
\mathcal{A}_{ABC\to ABC} = \sum_{i,j,k} \sum_{i',j',k'} \alpha_i \beta_j \gamma_k  \alpha_{i'} \beta_{j'}\gamma _{k'}   \mathcal{A}_{ijk \to i' j'k'}(p_1,p_2,p_3;p_4,p_5,p_6) \,,
\end{equation}
where the kinematics are chosen such that 
\begin{equation}
\mathcal{A}_{ijk \to i' j'k'}(p_1,p_2,p_3;p_4,p_5,p_6) = \mathcal{A}_{ijk \to i' j'k'}(p_1,p_2,p_3;-p_1,-p_2,-p_3)\,.
\end{equation}
Crossing symmetry for $3-3$ scattering for the chosen kinematics tells us that the amplitude is invariant under swapping $p_n$ with $p_{2n}=-p_n$. We can therefore write 
\begin{equation}
\mathcal{A}_{ABC\to ABC}^{\text{crossed}} = \sum_{i,j,k} \sum_{i',j',k'} \alpha_i \beta_j \gamma_k  \alpha_{i'} \beta_{j'}\gamma _{k'}  \mathcal{A}_{ijk' \to i' j'k}(p_1,p_2,p_3;-p_1,-p_2,-p_3) \,.
\end{equation}
Having already applied the forward limit, all that is left to do is to relabel dummy indices, and the result that follows is
\begin{equation}
\mathcal{A}_{ABC\to ABC}^{\text{crossed}} =\mathcal{A}_{ABC\to ABC}\,.
\end{equation}
A similar argument follows for even $n$. We have also explicitly checked the crossing symmetry of superposition amplitudes used in the results of this paper. 

\section{Symbolic Regression for Analytic Bounds}
\label{App:PySR}

Symbolic regression is a machine learning method for discovering mathematical expressions that describe the relationship between variables directly from data, without assuming a predetermined functional form. It is a supervised learning approach that searches over analytic expressions to find an optimised fit. This is opposed to other methods of fitting parameters with an already guessed model. It has been established as an important tool in the physical sciences \cite{Bartlett_2026}, proving useful in Beyond the Standard Model physics \cite{AbdusSalam:2024obf}, for example. We make use of the PySR tool for this \cite{cranmer2023interpretablemachinelearningscience}. \

\begin{figure}[h!]
    \centering
    \begin{subfigure}{1\textwidth}
        \centering
        \includegraphics[width=\textwidth]{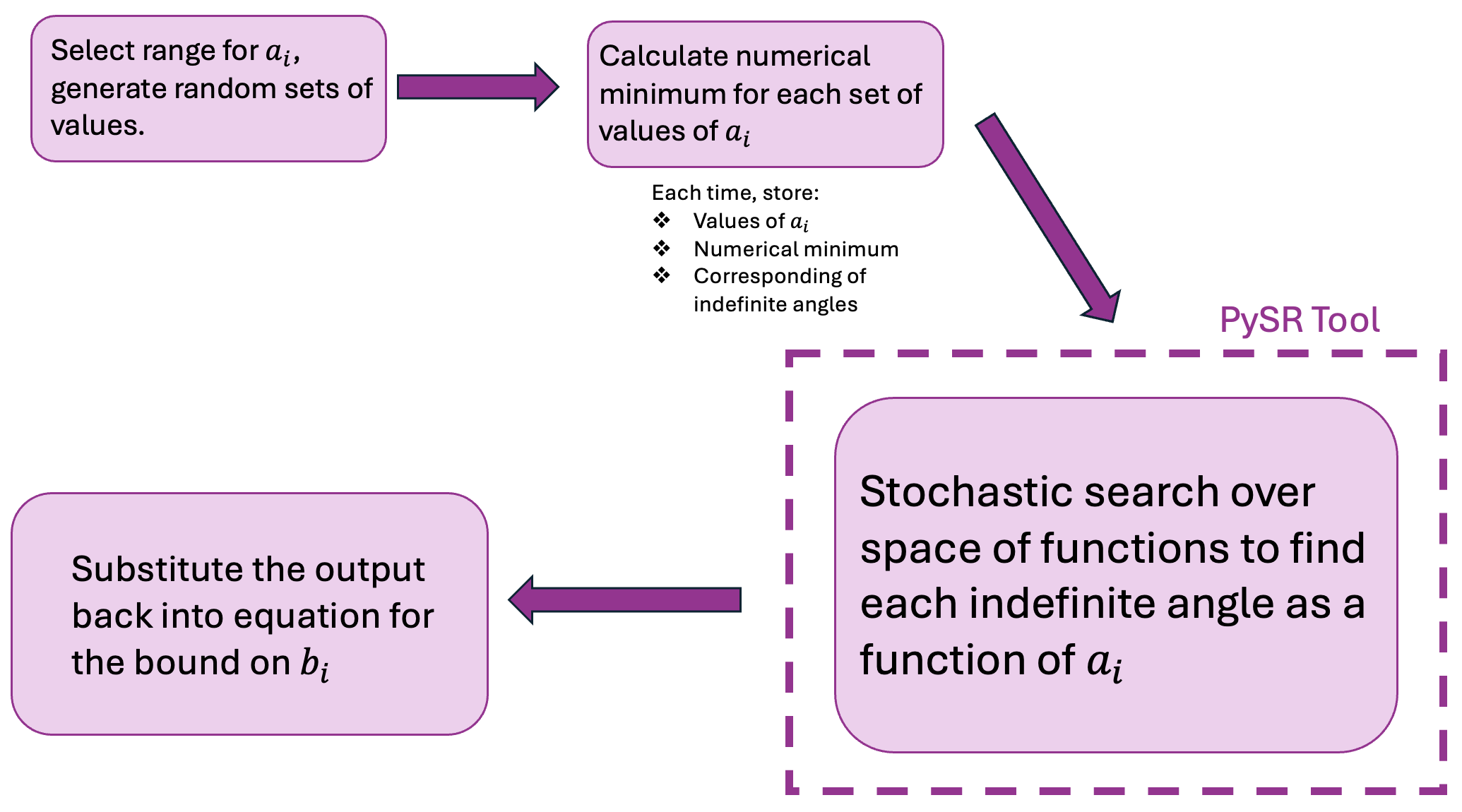}
        \caption{A flow diagram illustrating how we use symbolic regression to find analytic expressions for our bounds.}
    \end{subfigure}
    \vspace{0.5cm}
    \begin{subfigure}{1\textwidth}
        \centering
        \includegraphics[width=\textwidth]{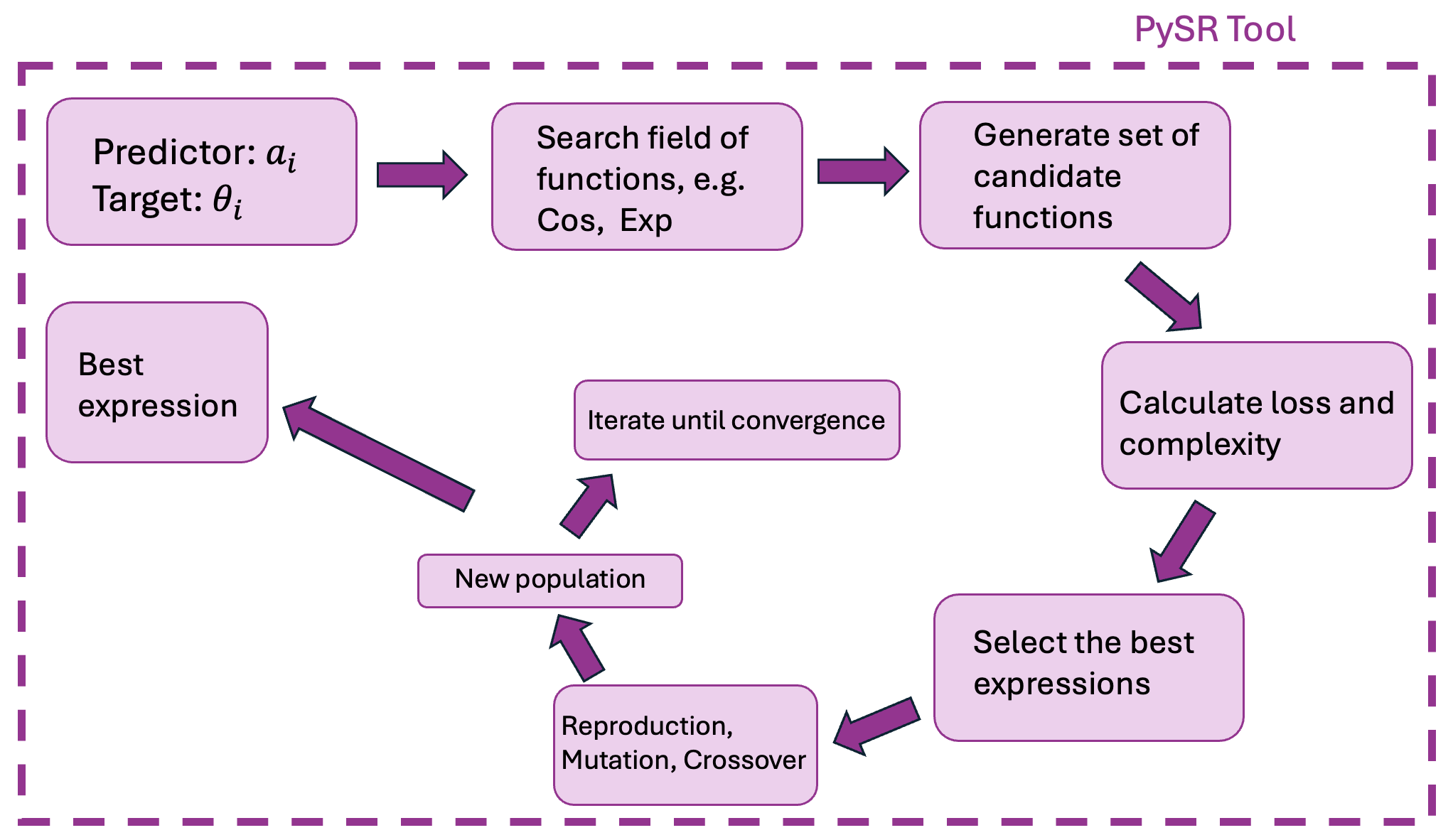}
        \caption{A flow diagram illustrating how symbolic regression works.}
    \end{subfigure}
    \caption{Two flow diagrams summarising the components of our computational method.}
    \label{fig:PySRworkflow1}
\end{figure}

Candidate expressions are represented as expression trees and evolved using a stochastic evolutionary algorithm. At each iteration, candidate expressions are evaluated according to their accuracy and complexity. Promising expressions are selected, and reproduction generates new expressions from these. Reproduction occurs primarily through Mutation, in which parts of an expression are randomly modified, and Crossover, in which parts of two parent expressions are combined to form a new expression. This process is repeated over many generations, producing expressions that provide a good balance between accuracy and simplicity. The resulting candidates form a set of solutions with the best trade off, from which simple and accurate symbolic expressions can be selected automatically. In general, the positivity bound will yield an expression to be satisfied which takes the form
\begin{equation}
    \sum_i A_i(\theta_j, \varphi_j)a_i + \sum_i B_i(\theta_j, \varphi_j)b_i >0 \quad \forall \theta_j, \varphi_j\,,
\end{equation}
where $a_i$ denote the Wilson coefficients which preserve the symmetry, and $b_i$ denote the ones that break it. We will always start by adding one symmetry breaking term at a time, so our bound becomes
\begin{equation}
     \sum_i A_i(\theta_j, \varphi_j)a_i +  B(\theta_j, \varphi_j)b >0 \quad \forall \theta_j, \varphi_j\,.
\end{equation}
When the function $B(\theta_j, \varphi_j)>0$, this captures the lower bound on $b$ in terms of the $a_i$. When $B(\theta_j, \varphi_j)<0$, this captures the upper bound on $b$. We apply this method to cases where $a_i>0$. We can therefore state that the strictest possible upper and lower bounds on $b$ come from
\begin{equation}
    b< \min_{\theta_j,\varphi_j: B(\theta_j, \varphi_j)<0} \left[\frac{A_i(\theta_j, \varphi_j)}{-B(\theta_j, \varphi_j)}\right]a_i \,,
\end{equation}
and 
\begin{equation}
    b> \max_{\theta_j,\varphi_j: B(\theta_j, \varphi_j)>0} \left[\frac{A_i(\theta_j, \varphi_j)}{-B(\theta_j, \varphi_j)}\right]a_i \,,
\end{equation}
and since we wish the bound to be satisfied for all values of the indefinite angles, we can combine these bounds even if the maxima and minima occur for different sets of angles. We are interested in finding these maxima and minima analytically. For a positivity bound to be valid, we must be able to trace back the indefinite state which produced it. In other words, the indefinite angles must be found as functions of the Wilson coefficients. Finding the correct functions is challenging with a large amount of parameters. This motivates our use of symbolic regression for this step. We choose ranges for the $a_i$ and calculate the numerical value of the upper and lower bound for each set of values of $a_i$. Each time, we store the value of the indefinite angles which produce the numerical minima. We loop over this procedure until we have collected around 1000 or more samples. Each sample contains the values of the $a_i$, the value of the numerical maxima and minima, and the set of indefinite angles which produced each. We then use this data for symbolic regression to find what each indefinite angle is in terms of the $a_i$. Once we obtain the best fits for all angles, we substitute this into the expressions for the bound and plot the allowed region alongside our numerical results from random angle sampling. The full process, including the logic of symbolic regression \cite{brum2026discoveringequationsdatasymbolic}, is summarised in Fig.~\ref{fig:PySRworkflow1}.
In addition, to improve the quality of the results, one can combine several different allowed bounds since the positivity bound must hold for all possible angle configurations.\ 

\begin{figure}[h!]
    \centering
    \includegraphics[width=0.9\linewidth]{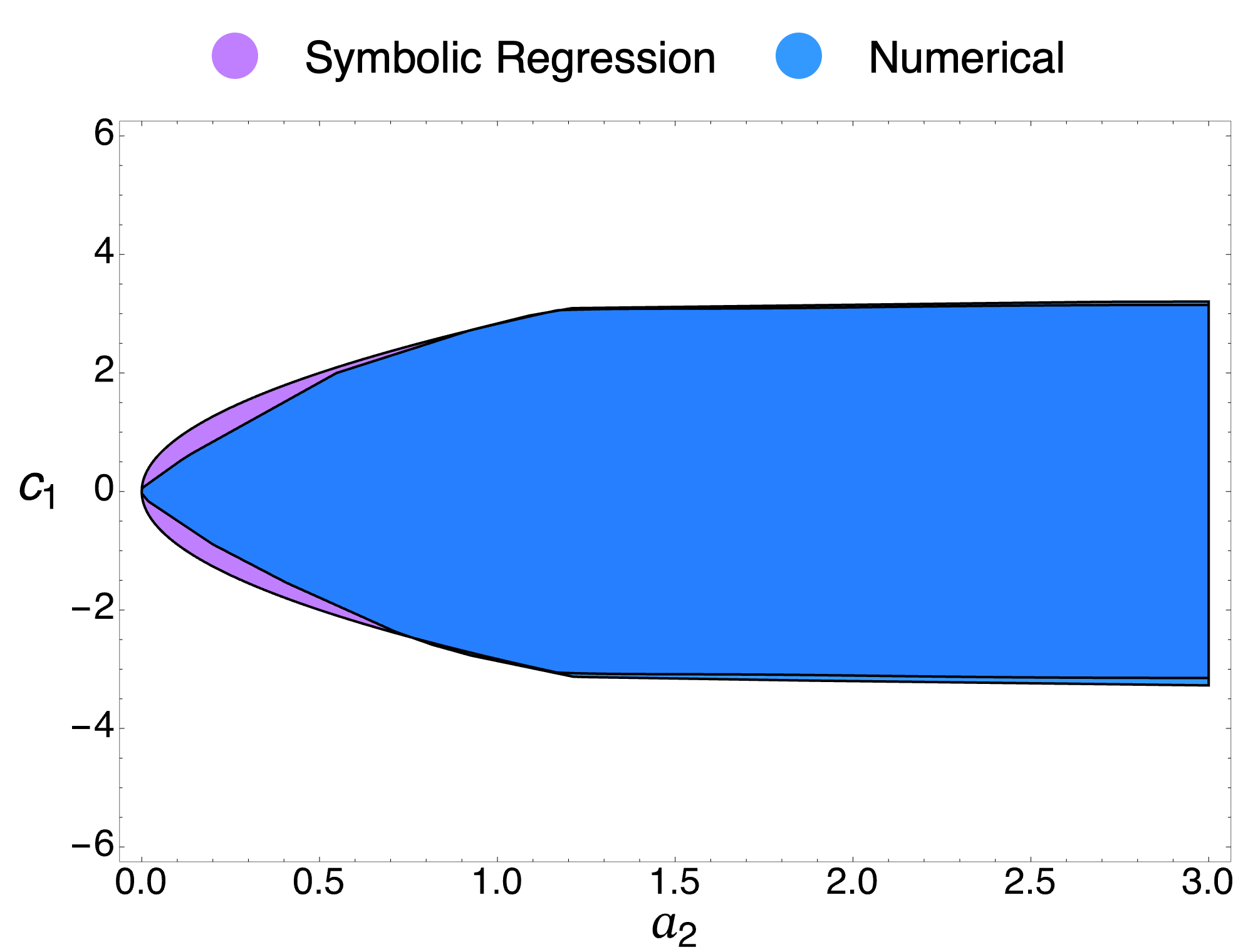}
    \caption{We can see that the expression recovered by the symbolic regression tool closely matches the numerical result. The result of the symbolic regression has been combined with the analytic bound.}
    \label{fig:2F4PnumericPySR}
\end{figure}

To illustrate the effectiveness of the symbolic regression method for recovering the analytic expression for the bound,
we look at the 2-field $2-2$ scattering in the case where  $a_1=1$ and $b_{1,2}=0=c_2$.  Fig.~\ref{fig:2F4PnumericPySR} compares the  numerical results of the two field example explored in Section~\ref{sec:Continuous} alongside the symbolic regression method.

\newpage
\bibliographystyle{JHEP}
\bibliography{biblio}

\end{document}